\documentclass[10pt,a4paper,final]{iopart}
\expandafter\let\csname equation*\endcsname\relax
\expandafter\let\csname endequation*\endcsname\relax
\usepackage[margin=1 in]{geometry}
\usepackage{amsmath,color,iopams,graphicx}
\usepackage[breaklinks=true,colorlinks=true,linkcolor=blue,urlcolor=blue,citecolor=blue]{hyperref}
\usepackage[english]{babel}
\usepackage{graphicx,comment}
\usepackage{mathtools}
\usepackage{amsmath}
\usepackage{amsfonts}
\usepackage{amssymb}
\usepackage{mathrsfs}
\usepackage{footnote}
\makesavenoteenv{tabular}
\makesavenoteenv{table}
\usepackage{multicols}
\usepackage{subfigure}
\usepackage{widetext}
\usepackage{caption}
\begin{document}
\title{Unconventional bosonization of chiral quantum wires coupled through a point-contact driven out of equilibrium}
\author{  Nikhil Danny Babu$^{\dagger}$, Girish S. Setlur$^{*}$}
\address{Department of Physics \\ Indian Institute of Technology  Guwahati \\ Guwahati, Assam 781039, India}
\ead{$^{\dagger}$danny@iitg.ac.in,$^{*}$gsetlur@iitg.ernet.in}

\begin{abstract}
Non-chiral bosonization technique (NCBT) adapted to study chiral quantum wires with non-interacting fermions coupled through a point-contact with a constant bias between the wires is introduced and is shown to reproduce the exact non-equilibrium Green functions of this system which was previously derived by the present authors analytically using standard methods. The tunneling I-V characteristics are obtained using the bosonized Green functions. The proposed unconventional bosonization scheme is also shown to be internally consistent as the four-point functions evaluated using NCBT are shown to be related to the two point functions through Wick's theorem as it should be. In equilibrium, the equal space-time NCBT Green functions for an interacting Luttinger liquid with impurities obtained in a previous work shows universal scaling behaviour in accordance with Bethe ansatz and functional renormalization group predictions. We expect to obtain a similar universal scaling form of the tunneling properties out of equilibrium in presence of interparticle interactions.
\end{abstract}

\vspace{2pc}
\section{INTRODUCTION}
The quantum physics of low-dimensional systems has been a source of rich theoretical results in condensed matter physics for the past few decades. The advent of modern experimental techniques has made it possible to investigate such systems in practice in the  context of nanostructures, carbon nanotubes, quantum Hall edges and confined ultracold atoms. In one-dimension, interacting fermionic systems are described by the Luttinger liquid paradigm \cite{stone1994bosonization,Delft,Haldane1981,giamarchi2003quantum}. Since the early works of Tomonaga \cite{10.1143/ptp/5.4.544} and Luttinger \cite{doi:10.1063/1.1704046} it has been evident that even at low energies the interparticle interactions cannot be treated perturbatively in Luttinger liquids in contrast to the case of Fermi liquids in higher dimensions. The technique of bosonization was developed to deal with interactions in one-dimensional systems and it has been successful to a large extent. Bosonization involves replacing fermionic excitations with bosonic degrees of freedom. This technique is widely used and is well understood in the context of one-dimensional systems in equilibrium. Recently there is a growing interest in nonequilibrium phenomena in nanostructures and quantum wires \cite{PhysRevLett.102.036804,doi:10.1063/1.3253705,altimiras2010non}. Bosonization is a very versatile technique and a major fraction of the theoretical research on one-dimensional systems relies on this technique which could, in principle be extended to out-of-equilibrium situations as well. It has been well demonstrated in the literature that a consistent bosonization scheme could prove to be a powerful tool in analysing nonequilibrium transport problems in one-dimension.
Using the framework of the Keldysh action formalism Gutman et.al \cite{PhysRevB.81.085436,Gutman_2010} have developed a bosonization technique for one-dimensional fermions in nonequilibrium and subsequently used this method to study an interacting quantum wire attached to two electrodes with arbitrary energy distributions but their methods work with the crucial assumption of no electron backscattering due to impurities. Conventional or standard bosonization is ill-suited to study systems where there is backscattering from impurities. In this article we restrict to leads with noninteracting chiral fermions in contact with reservoirs at different chemical potentials and with a point contact tunneling junction at the origin \cite{RevModPhys.71.S306,BLANTER20001}. This is similar to the situation considered in \cite{PhysRevB.93.085440}, where it was shown that the conventional bosonization-debosonization procedure fails to give results that are consistent with the exact solution to this noninteracting problem. In this paper we set up a bosonization formalism for this problem in the spirit of the \textit{Non-chiral bosonization technique (NCBT)} \cite{doi:10.1142/S0217751X18501749} which involves a radical new way of looking at the Fermi-Bose correspondence as a mnemonic to obtain the correct correlation functions rather than an operator identity and has been proven to be more successful than conventional bosonization in studying Luttinger liquids with impurities \cite{das2019nonchiral,doi:10.1142/S0217751X18501749,DAS2017216,Das_2018,DAS201939,DAS20193149,Das_2020,Danny_Babu_2020}. The present work extends this idea to systems that are driven out of equilibrium by the application of a bias between the left and right movers. We show that an appropriate version of NCBT reproduces the exact single-particle Green functions of this system under bias. \\ \mbox{ }\\
The paper is organized as follows. In Sec.\ref{secmod} we introduce the model and its exact Green functions. In Sec.\ref{secdensity} we discuss the density-density correlation functions in the presence of a constant bias. In Sec.\ref{secbos} we explain the unconventional method of bosonization we employ which involves a modification of the standard Fermi-Bose correspondence and we show how to obtain the noninteracting two-point correlation functions using this bosonization scheme. Sec.\ref{fourpoint} is concerned with  the evaluation of relevant four-point functions using this unconventional bosonization procedure and this is a crucial cross-check that validates our method especially when we show that these results are consistent with Wick's theorem. In Sec.\ref{currentcond} we obtain the known results for the tunneling current and conductance from the bosonized Green functions. Conclusions and a brief summary of the future prospects for this work is presented in Sec.\ref{conclusion}.
 \section{MODEL DESCRIPTION}
 \label{secmod}
 We consider the same model as we did in our previous work \cite{Babu_2022} wherein we have computed the exact space-time nonequilibrium Green function (NEGF) for the problem. The model Hamiltonian is
 \begin{align}
H = \sum_{p}(v_Fp+e V_{b}(t) )c^{\dagger}_{p,R}c_{p,R} &+ \sum_{p}(-v_Fp)c^{\dagger}_{p,L}c_{p,L}+ \frac{\Gamma}{L}(c^{\dagger}_{.,R}c_{.,L}+c^{\dagger}_{.,L}c_{.,R})
 \label{eqham}
\end{align}
where $R$ and $L$ label the right and left moving chiral spinless modes and $V_{b}(t)$ is the generic time dependent bias voltage applied to one of the contacts , the right($R$) moving one in this case. $c^{\dagger}_{p}$ and $c_{p}$ are the spinless fermion creation and annihilation operators in momentum space and we use the notation $c^{\dagger}_{.,R} = \sum_{p}c^{\dagger}_{p,R}$ and $\Gamma = \Gamma^{*}$ is the tunneling amplitude of the symmetric point-contact junction and $L$ that does not appear in the subscript is the system size. We have considered a generic time-dependent bias potential defined as $\mu_{L}-\mu_{R} = -\mu_{R} = -e V_{b}(t) = e V(t)$ following the same convention as in \cite{PhysRevB.93.085440}. Note that we assume the bias to be applied only to the right movers, hence $\mu_{L} = 0$ and $\mu_{R} = e V_{b}(t)$. In \cite{Babu_2022} we obtain the exact dynamical nonequilibrium Green function (NEGF) by solving Dyson's equation analytically and this allows us to write down the two-point and four-point functions for this system in a closed form in terms of simple functions of position and time. The full NEGF as obtained in \cite{Babu_2022} is,
\begin{align}
 <\psi^{\dagger}_{\nu^{'}}(x^{'},t^{'})\psi_{\nu}(x,t)> \mbox{ }=\mbox{ } -\frac{i}{2\pi} \frac{\frac{\pi}{\beta v_{F}}}{\sinh( \frac{ \pi }{\beta v_F } (\nu x-\nu^{'}x^{'}-v_F(t-t^{'}) ) )} \kappa_{\nu,\nu^{'}}
\label{eqnoneq}
 \end{align}
 where $\nu$,$\nu^{'} = \pm 1$ with $R=1$ and $L=-1$ and
 \begin{widetext}
 \small
 \begin{align}
 \kappa_{1,1} \mbox{ }=\mbox{ } \bigg(  U(t^{'},t) \left[1
-      \theta(x^{'})    \mbox{          }   \frac{  2 \Gamma^2
}{\Gamma ^2 +4 v_F^2}\right]&\mbox{          }    \left[1
-      \theta(x )    \mbox{          }   \frac{  2 \Gamma^2
}{\Gamma ^2 +4 v_F^2}\right]
  \nonumber \\ &+ \left( \frac{\Gamma }{v_F} \mbox{          } \frac{(2 v_F)^2
}{\Gamma ^2 +4 v_F^2}\right)^2 \mbox{          }
  \theta(x )     \theta(x^{'} )   \mbox{          } U(t^{'},t^{'} - \frac{x^{'}}{v_F})   \mbox{     }
   U(t - \frac{x}{v_F},t)   \bigg)
   \label{eqkappa1}
 \end{align}
 \begin{align}
 \kappa_{-1,-1} \mbox{ }=\mbox{ } \bigg( \left[ 1
-      \theta(-x^{'})    \mbox{          }   \frac{  2 \Gamma^2
}{\Gamma ^2 +4 v_F^2}
\right] &  \left[ 1
-      \theta(-x )    \mbox{          }   \frac{  2 \Gamma^2
}{\Gamma ^2 +4 v_F^2}
\right]   \nonumber \\ &+   \left( \frac{\Gamma }{v_F} \mbox{          } \frac{(2 v_F)^2
}{\Gamma ^2 +4 v_F^2}\right)^2 \mbox{          }  \theta(-x )    \theta( -x^{'} )    \mbox{          }
 U(t^{'} + \frac{x^{'}}{v_F},t + \frac{x}{v_F} ) \bigg)
 \label{eqkappa2}
 \end{align}
 \begin{align}
 \kappa_{1,-1} \mbox{ }= \mbox{ }\bigg( -U(t - \frac{x}{v_F},t)   \mbox{          }  &\left[ 1
-      \theta(-x^{'})    \mbox{          }   \frac{  2 \Gamma^2
}{\Gamma ^2 +4 v_F^2}
\right]
    \theta(x )   \mbox{          } \nonumber \\ &+
 U(t^{'} + \frac{x^{'}}{v_F},t)
\mbox{          }
\left[1
-      \theta(x )    \mbox{          }   \frac{  2 \Gamma^2
}{\Gamma ^2 +4 v_F^2}\right] \theta( -x^{'} )   \bigg) i  \frac{\Gamma }{v_F}\frac{(2 v_F)^2
}{\Gamma ^2 +4 v_F^2}
\label{eqkappa3}
 \end{align}
 \begin{align}
 \kappa_{-1,1} \mbox{ }=\mbox{ }\bigg( - U(t^{'},t + \frac{x}{v_F} ) &\left[1
-      \theta(x^{'})    \mbox{          }   \frac{  2 \Gamma^2
}{\Gamma ^2 +4 v_F^2}\right]  \theta( -x )
 \nonumber \\ &+     U(t^{'},t^{'} - \frac{x^{'}}{v_F})   \mbox{          }     \left[ 1
-      \theta(-x )    \mbox{          }   \frac{  2 \Gamma^2
}{\Gamma ^2 +4 v_F^2}
\right]  \theta(x^{'} )     \bigg) \mbox{          } i \frac{\Gamma }{v_F}    \frac{ (  2  v_F )^2   }{\Gamma ^2 +4 v_F^2}
\label{eqkappa4}
 \end{align}
 \end{widetext}
 Here $U(\tau,t) \equiv e^{ - i \int _{ \tau }^t  e V_{b}(s)ds }$ and $\theta(x)$ is the Dirichlet regularized step function (this means $ \theta(x > 0) = 1, \theta(x < 0) = 0, \theta(0) = \frac{1}{2} $). In the equilibrium limit (limit of zero bias) i.e. $U(\tau,t) \equiv 1$, the NEGF in the above equation reduces to a similar form as the equilibrium noninteracting Green functions in \cite{doi:10.1142/S0217751X18501749} but they are not fully the same since in \cite{doi:10.1142/S0217751X18501749} the impurity terms in the Hamiltonian include both forward scattering and backward scattering terms, however in our present case only backward scattering (tunneling term) from the impurity (point-contact) is considered. Upon sudden switch on of a bias, the NEGF shows an initial transient regime before the appearance of a steady state. However, the transport characteristics show only steady state behaviour (at least in the case of a point contact with infinite bandwidth in the momentum to which we will be restricting ourselves in the rest of the paper). To develop a bosonization scheme that can reproduce the nonequilibrium Green functions is a challenging task even in the absence of interparticle interactions.  The Fermi-Bose correspondence of chiral fermions is an exact operator identity for chiral fermions \footnote{The full proof of this, especially the proof that the conventional fermion properly anticommutes with its bosonized version, is not available in the literature but was proved by F.D.M. Haldane in a private communication with one of the authors and it was subsequently made available in a series of lectures online (\textit{https://www.youtube.com/watch?v=dX17Xe5OZh0\&t=1532s})}. But the presence of a point-contact in the model breaks translation invariance and hence conventional bosonization formulas are unwieldy in this case since the framework has been developed keeping the free particle in mind (e.g. all momentum states upto some level are filled for each chirality suggesting free particles and the number of each chirality is conserved). But in case of a half-line, it is possible to adapt this approach to even account for mutual interactions between fermions. Such situations have been dealt with in \cite{PhysRevB.83.085415} by bosonizing a translation invariant reference system and imposing appropriate boundary conditions.\\
 But in order to obtain the full space-time correlation functions in a closed form for an impurity of a finite strength (not vanishingly small or infinitely large), the use of radical new ideas like the non-chiral bosonization technique (NCBT) becomes important.  In the NCBT approach, the notion that the Fermi-Bose correspondence is a strict operator identity is eased and the point of view taken is that the Fermi-Bose correspondence is a mnemonic valid at the level of correlation functions. This technique has been successfully used to obtain the most singular parts of the correlation functions in strongly inhomogeneous Luttinger liquids in equilibrium \cite{doi:10.1142/S0217751X18501749}. The effect of the backward scattering of the fermions due to an impurity is encoded in the Fermi-Bose correspondence and the Hamiltonian remains local in this method (unlike conventional approaches where the Fermi-Bose correspondence is sacred, as a result the Hamiltonian is nonlinear and nonlocal when impurities are present).
 \section{DENSITY-DENSITY CORRELATION FUNCTIONS WITH A CONSTANT BIAS}
 \label{secdensity}
  The density-density correlation functions (DDCFs) are certain four-point functions that are crucial to this approach. In the absence of interparticle interactions and assuming a constant bias one can use Wick's theorem to write the DDCF as follows,
  \begin{align}
  <T \mbox{ }\rho_{\nu}(x,t)\rho_{\nu^{'}}(x^{'},t^{'})>_{0} - <\rho_{\nu}(x,t)>_{0}<\rho_{\nu^{'}}(x^{'},t^{'})>_{0} \nonumber \\\mbox{ }=\mbox{ } - <T \mbox{ }\psi_{\nu}(x,t) \psi^{\dagger}_{\nu^{'}}(x^{'},t^{'})>_{0}\mbox{ } <T \mbox{ }\psi_{\nu^{'}}(x^{'},t^{'}) \psi^{\dagger}_{\nu}(x,t) >_{0}
  \end{align}
  Here the subscript $<\mbox{ }\mbox{ }>_{0}$ denotes the absence of mutual interactions. Assuming a constant bias we get the following expressions for the DDCF:
 \begin{widetext}
 \begin{align}
  <&T \mbox{     } \rho_R(x,t) \rho_R(x^{'},t^{'}) >_0 -<\rho_R(x,t) >_0<\rho_R(x^{'},t^{'}) >_0\mbox{          }
 \nonumber \\ &=  \mbox{          }<T \mbox{     } \rho_L(-x,t) \rho_L(-x^{'},t^{'}) >_0-<\rho_L(-x,t) >_0<\rho_L(-x^{'},t^{'}) >_0 \mbox{          }\nonumber \\
&=\mbox{ } \left[    \frac{i}{2\pi} \frac{ \frac{ \pi }{\beta v_F } }{\sinh( \frac{ \pi }{\beta v_F } (x-x^{'}-v_F(t-t^{'}) ) )  } \right]^2 \nonumber \\ &
\mbox{          }\mbox{          }\mbox{          }\mbox{          }\mbox{          }
    \bigg(   \left[1
-      \theta(x^{'})    \mbox{          }   \frac{  2 \Gamma^2
}{\Gamma ^2 +4 v_F^2}\right]^2\mbox{          }    \left[1
-      \theta(x )    \mbox{          }   \frac{  2 \Gamma^2
}{\Gamma ^2 +4 v_F^2}\right]^2
  + \left( \frac{\Gamma }{v_F} \mbox{          } \frac{(2 v_F)^2
}{\Gamma ^2 +4 v_F^2}\right)^4 \mbox{          }
  \theta(x )     \theta(x^{'} )  \nonumber \\
  &\mbox{ }\mbox{          }\mbox{          }\mbox{          }+
\mbox{          }
\left[1
-       \frac{  2 \Gamma^2
}{\Gamma ^2 +4 v_F^2}\right]^2 \mbox{          }   \left( \frac{\Gamma }{v_F} \mbox{          } \frac{(2 v_F)^2
}{\Gamma ^2 +4 v_F^2}\right)^2 \mbox{          } \theta(x )     \theta(x^{'} )   \mbox{          }
   \left(   e^{ -i e V_b (t- t^{'} - \frac{x}{v_F} + \frac{x^{'}}{v_F}) }  +   e^{ -i e V_b (t^{'}-t  - \frac{x^{'}}{v_F}+ \frac{x}{v_F}) }
  \right)\bigg)
  \label{rhorhoRRLL}
 \end{align}
 \begin{align}
 <&T \mbox{     } \rho_R(x,t) \rho_L(-x^{'},t^{'}) >_0-<\rho_R(x,t)>_0< \rho_L(-x^{'},t^{'}) >_0 \mbox{          }
 \nonumber \\ &=  \mbox{          }<T \mbox{     } \rho_L(-x,t) \rho_R(x^{'},t^{'}) >_0 - <\rho_L(-x,t) >_0<\rho_R(x^{'},t^{'}) >_0\mbox{          }\nonumber \\
 &= \mbox{ }\left[  \frac{i}{2\pi} \frac{ \frac{ \pi }{\beta v_F } }{\sinh( \frac{ \pi }{\beta v_F } (x-x^{'}-v_F(t-t^{'}) ) )  }\right]^2 \nonumber \\ &
 \mbox{          }\mbox{          }\mbox{          }\mbox{          }\mbox{          }
 \bigg(\left( i  \frac{\Gamma }{v_F}\frac{(2 v_F)^2
}{\Gamma ^2 +4 v_F^2} \right)^2
\mbox{          }
  \left( -    \left[ 1
-      \theta(x )    \mbox{          }   \frac{  2 \Gamma^2
}{\Gamma ^2 +4 v_F^2}
\right]^2    \theta(x^{'})  -
 \left[1
-      \theta(x^{'} )    \mbox{          }   \frac{  2 \Gamma^2
}{\Gamma ^2 +4 v_F^2}\right]^2    \theta( x)      \right) \nonumber \\
&\mbox{          }\mbox{          }\mbox{          }\mbox{          }+  \left( i  \frac{\Gamma }{v_F}\frac{(2 v_F)^2
}{\Gamma ^2 +4 v_F^2} \right)^2
\mbox{          }
  \left(e^{ -i e V_b (t^{'}-t - \frac{x^{'}}{v_F} + \frac{x}{v_F}) }    +e^{ -i e V_b (t-t^{'} - \frac{x}{v_F} + \frac{x^{'}}{v_F}) }   \right)   \mbox{          }     \left[ 1
-       \frac{  2 \Gamma^2
}{\Gamma ^2 +4 v_F^2}
\right]^2  \theta(x)
  \theta( x^{'} )\bigg)
  \label{rhorhoRLLR}
 \end{align}
 \end{widetext}
 It is easy to check using the above equations that the density-density correlation functions satisfy the following identities,
 \begin{align}
 <&T \mbox{     } \rho_{\nu}(\nu \mbox{ }x,t) (\rho_{\nu}(\nu\mbox{ }x^{'},t^{'})+\rho_{-\nu}(-\nu \mbox{ }x^{'},t^{'})) >_0 -<\rho_{\nu}(x,t) >_0<(\rho_{\nu}(\nu \mbox{ }x^{'},t^{'})+\rho_{-\nu}(-\nu\mbox{ }x^{'},t^{'})) >_0\mbox{          }=
 \mbox{          }\nonumber \\
  <&T \mbox{     }(\rho_{\nu}(\nu \mbox{ }x,t) + \rho_{-\nu}(-\nu \mbox{ }x,t)  ) \rho_{\nu}(\nu \mbox{ }x^{'},t^{'}) >_0 - <(\rho_{\nu}(\nu \mbox{ }x,t) + \rho_{-\nu}(-\nu \mbox{ }x,t) ) >_0<\rho_{\nu}(\nu\mbox{ }x^{'},t^{'}) >_0 \nonumber \\ &\mbox{     }=  \mbox{          }
  \left[    \frac{i}{2\pi} \frac{ \frac{ \pi }{\beta v_F } }{\sinh( \frac{ \pi }{\beta v_F } (x-x^{'}-v_F(t-t^{'}) ) )  } \right]^2
 \end{align}
 where $\nu$ take values $\pm 1$ with $1$ denoting $R$ (right movers) and $-1$ denoting $L$ (left movers).
 \section{BOSONIZATION SCHEME}
 \label{secbos}
 \subsection{General formalism}
 Because the fermions have a linear dispersion, it is expected that one should be able to express the low-energy degrees of freedom in terms of bosons. The Fourier transform of the right mover density is,
 \begin{align}
 \rho_{R}(x,t) = \frac{1}{L}\sum_{q} \rho^{R}_{q} e^{-i q x}
 \end{align}
 where $\rho^{R}_{q} = \sum_{k} :c^{\dagger}_{k,R} c_{k+q,R} :$ . The commutator of the densities is
 \begin{align}
 [\rho^{R}_{q},\rho^{R}_{q^{'}}]  
 &= \frac{q L}{2 \pi}\delta_{q+q^{'},0}
 \end{align} 
  A similar calculation gives $[\rho^{L}_{q},\rho^{L}_{q^{'}}] = -\frac{q L}{2 \pi}\delta_{q+q^{'},0}$, hence apart from a numerical factor the density operators are bosons.
 In conventional bosonization, the chiral Fermi fields are expressed in terms of the density as,
 \begin{align}
 &\psi_{\nu}(x,t) = e^{ 2 \pi i \mbox{ }\nu\int^{x} \rho_{\nu}(y,t) dy }
 \label{standardbose}
 \end{align}
 where $\nu = \pm$ for $R$ or $L$ respectively. The density is related to bosonic fields $\phi_{\nu}(x,t)$ as
 \begin{align}
 2 \pi \int^{x} dy \mbox{ } \rho_{\nu}(y,t) = \phi_{\nu}(x,t)
 \end{align}
 and they obey the commutation relations
 \begin{align}
 [\phi_{R}(x,t), \phi_{R}(x^{'},t)] = -[\phi_{L}(x,t), \phi_{L}(x^{'},t)] = -i \pi \mbox{ }  sgn(x-x^{'})
 \end{align}
 The fermionic anticommutation relations hold in this boson representation \cite{Emery1979}. The Fermi-Bose correspondence in Eq. \ref{standardbose} is  proved by resorting to the basis states of homogeneous (translation-invariant) systems. When we have an impurity in the system like the point-contact tunnel junction present in the model under consideration, the number of right movers and left movers are not separately conserved, hence bosonization using the conventional approach is very inconvenient.
\subsection{Non-chiral bosonization}
 We handle this situation by modifying the standard Fermi-Bose correspondence to include the effects of backscattering of fermions from the impurity. The modified Fermi-Bose correspondence takes the form,
 \begin{align}
 \psi_{\nu}(x,t) = e^{ 2 \pi i \nu  \int^{x} (\rho_\nu(y,t)+\lambda \mbox{ }\rho_{-\nu}(-y,t)) dy }
 \label{ncbt}
 \end{align}
 where the value of $\lambda$ which is either $0$ or $1$ dictates the presence of the additional $\rho_{-\nu}(-y,t)$ term. The presence of a bias completely spoils the partial symmetry that previously existed between left and right movers. This necessitates a completely new ansatz to replace the original ansatz found in previous works published on the non-chiral bosonization technique, although superficially they look similar. There is no obvious reason why any of these methods ought to work beforehand. It is only after we try out various possibilities that we come to the conclusion that the present approach is the only one that works. This is especially true for example when we find that two different ansatzs reproduce a certain two-point function, but only one of them gives the correct four-point function thus ruling out the other (please refer to Sec.\ref{fourpoint}). We demand that the correlation functions calculated using the correspondence in Eq. \ref{ncbt} reproduce the exact Green functions in Eq. \ref{eqnoneq}.

   Since $[\rho_{R}(x),\rho_{L}(x^{'})] = 0$ and $[\rho_{\nu}(x),\rho_{\nu}(x^{'})] = 
   \nu \frac{i}{2 \pi}\delta^{'}(x-x^{'})$, fermion anticommutation relations are satisfied when $x \neq x^{'}$ at the level of operators even for this modified Fermi-Bose correspondence seen in Eq.(\ref{ncbt}). For $x = x^{'}$, however, the correct anticommutation relations are only recovered at the level of correlation functions. One should note that the NCBT ansatz in Eq.\ref{ncbt} is not a strict operator identity but merely a mnemonic to get the correlation functions. It has been shown in previous works that the non-chiral bosonization technique (NCBT) can be used to obtain the most singular parts of the Green's functions of interacting Luttinger liquids with impurities exactly \cite{das2019nonchiral}.  The series expansion of the NCBT Green's functions in powers of fermion-fermion interaction strength matches term by term with standard fermionic perturbation theory (most singular terms). It has also been shown that the NCBT Green's functions with forward scattering between fermions satisfy the (most singular parts of the) exact Schwinger-Dyson equations \cite{das2019nonchiral}. These results settle any suspicion the reader may have that this formalism is mere phenomenology.

 There are several terms that can possibly contribute to the evaluation of the two-point function. These are listed below:
 \begin{widetext}
 \begin{align}
 <&T\mbox{ }\psi_{\nu}(x,t) \psi^{\dagger}_{\nu^{'}}(x^{'},t^{'})> \mbox{ }=\mbox{ } < e^{ 2 \pi i \nu \int^{x} \rho_\nu (y,t) \mbox{      }dy }\mbox{ } e^{ -2 \pi i \nu^{'} \int^{x^{'}} \rho_{\nu^{'}}(y^{'},t^{'}) \mbox{      }dy^{'} }  >  \label{vt1}
 \\
  <&T\mbox{ }\psi_{\nu}(x,t) \psi^{\dagger}_{\nu^{'}}(x^{'},t^{'})> \mbox{ }=\mbox{ } <e^{ 2 \pi i \nu \int^{x} (\rho_\nu (y,t) + \rho_{-\nu}(-y,t) )  \mbox{      }dy } \mbox{ } e^{ -2 \pi i \nu^{'}\int^{x^{'}} \rho_{\nu^{'}}(y^{'},t^{'}) \mbox{      }dy^{'} } >  \label{vt2}\\
  <&T\mbox{ }\psi_{\nu}(x,t) \psi^{\dagger}_{\nu^{'}}(x^{'},t^{'})> \mbox{ }=\mbox{ }<e^{ 2 \pi i \nu \int^{x} \rho_\nu(y,t) \mbox{      }dy }\mbox{ } \mbox{ } e^{ -2 \pi i \nu^{'}\int^{x^{'}} (\rho_{\nu^{'}}(y^{'},t^{'}) + \rho_{-\nu^{'}}(-y^{'},t^{'}) )  \mbox{      }dy^{'} }>  \label{vt3}\\
  <&T\mbox{ }\psi_{\nu}(x,t) \psi^{\dagger}_{\nu^{'}}(x^{'},t^{'})> \mbox{ }=\mbox{ }<e^{ 2 \pi i \nu \int^{x} (\rho_\nu(y,t) + \rho_{-\nu}(-y,t) )  \mbox{      }dy } \mbox{ }  \mbox{ } e^{ -2 \pi i \nu^{'} \int^{x^{'}} (\rho_{\nu^{'}}(y^{'},t^{'}) + \rho_{-\nu^{'}}(-y^{'},t^{'}) ) dy^{'} }> \label{vt4}
 \end{align}
 \end{widetext}
These terms are evaluated using a version of the cumulant expansion (Baker-Campbell-Haussdorff formula).
\begin{align}
<e^{A} e^{B}> \sim e^{\frac{1}{2}<A^{2}>} e^{\frac{1}{2}<B^{2}>} e^{<AB>}
\label{bch}
\end{align}
 The terms of the type in Eq.\ref{vt4} can be discarded straightaway as they do not even correctly reproduce the $\sinh( \frac{ \pi }{\beta v_F } (\nu x-\nu^{'}x^{'}-v_F(t-t^{'}) ) )^{-1} $ factor that appears in the Green function, instead they give the square of that term. A linear combination of the other three terms reproduce the exact Green functions apart from the bias dependent exponential prefactors $U(\tau,t) = e^{-i \int_{\tau}^{t} e V_{b} dt}$. However we note that terms of the standard bosonization form $< e^{ 2 \pi i \nu \int^{x} \rho_\nu (y,t) \mbox{      }dy }\mbox{ } e^{ -2 \pi i \nu^{'} \int^{x^{'}} \rho_{\nu^{'}}(y^{'},t^{'}) \mbox{      }dy^{'} }  > $ do not always reproduce the correct form of correlation functions. For example let us evaluate for $x>0, x^{'}>0$,
\begin{align}
< e^{ 2 \pi i  \int^{x} \rho_R (y,t) \mbox{      }dy }\mbox{ } e^{ -2 \pi i  \int^{x^{'}} \rho_{R}(y^{'},t^{'}) \mbox{      }dy^{'} }  > \mbox{ }=\mbox{ }&
e^{\frac{1}{2} (2 \pi i)^2  \int^{x}  \mbox{      }dy \mbox{         } \int^{x}  \mbox{      }dy^{'} \mbox{         } <\rho_R(y,t)\rho_R(y^{'},t) > }\mbox{ }\nonumber \\ &e^{ \frac{1}{2} (2 \pi i)^2  \int^{x^{'}}  \mbox{      }dy \int^{x^{'}}  \mbox{      }dy^{'} \mbox{      } < \rho_R(y,t^{'})\rho_R(y^{'},t^{'}) > } \nonumber \\   & e^{ -(2 \pi i)^2 \int^{x}  \mbox{      }dy \int^{x^{'}}  \mbox{      }dy^{'} \mbox{      } < \rho_R(y,t)\rho_R(y^{'},t^{'})> }
\end{align}
The rhs of the above equation upon evaluation for $x>0, x^{'}>0$ using Eq.\ref{rhorhoRRLL} doesn't reproduce the dynamical part of the exact $RR$ Green function. The conventional bosonization choice in Eq.\ref{vt1} only works for the cases: $<\psi_{R}(x<0,t) \psi^{\dagger}_{R}(x^{'}<0,t^{'})>$ and $<\psi_{L}(x>0,t) \psi^{\dagger}_{L}(x^{'}>0,t^{'})>$. We show in Section.\ref{fourpoint} where we evaluate the four-point functions that for these two cases the conventional choice that amounts to setting $\lambda=0$ in Eq.\ref{ncbt} is the one that works. As for all other cases it is necessary to use the modified Fermi-Bose correspondence i.e. $\lambda=1$ in Eq.\ref{ncbt}, this means that in order to obtain the two-point functions in these cases the Eqs.\ref{vt2} and \ref{vt3} must be used.\\
 In order to obtain the correct bias dependent $U(\tau,t)$ terms as in Eq.\ref{eqnoneq} using this bosonization scheme, we make the following unitary field transformations
 \begin{align}
 &\tilde{\psi}_R(x,t) \mbox{ }=\mbox{ } U(-\infty,t)\psi_{R}(x,t)  \\
 &\hat{\psi}_R(x,t)\mbox{ }=\mbox{ } U(t-\frac{x}{v_{F}},t)\psi_{R}(x,t) \\
 &\tilde{\psi}_L(x,t) \mbox{ }=\mbox{ }\psi_{L}(x,t) \\
 &\hat{\psi}_{L}(x,t) \mbox{ }=\mbox{ }U(-\infty, t + \frac{x}{v_{F}})\psi_{L}(x,t)
 \end{align}
\begin{figure}
\centering
 \includegraphics[scale=0.09]{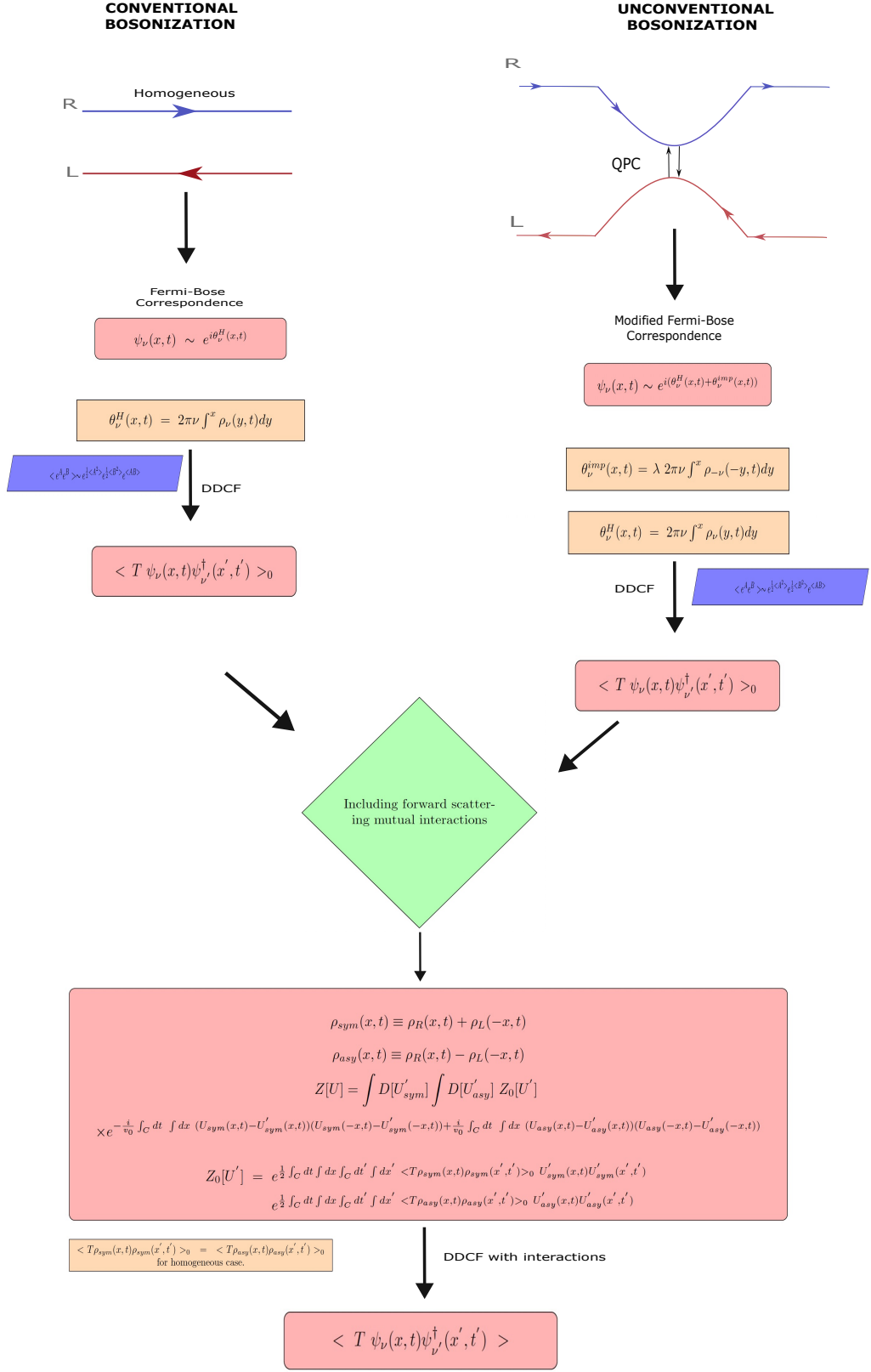}
 \caption{\small Flowchart diagram highlighting the main differences between the standard bosonization procedure and the NCBT based unconventional bosonization scheme we propose for an inhomogeneous system with a point-contact. Here $\nu, \nu^{'} = \pm 1$ denoting $R(+1)$ and $L(-1)$. The parameter $\lambda = 0,1$ denoting the absence or presence of the unconventional term in the bosonization formula and $v_{0}$ is the strength of the mutual interaction between the right and left movers.}
 \label{fig1}
 \end{figure}
This asymmetrical transformations are invoked because the bias is only applied to one of the chiralities. We construct a bosonization prescription using a combination of the above mentioned field transformations along with the bosonized version of the two-point functions in Eqs.\ref{vt1},\ref{vt2} and \ref{vt3} that reproduces the full non-interacting Green functions apart from constant prefactors. The correlations of the transformed fields can be written in a bosonized form as follows.\\
For the $RR$ case,
\begin{align}
< T\mbox{ }\hat{\psi}_R(x,t) \hat{\psi}^{\dagger}_R(x^{'},t^{'})>_{\lambda=1}&= U(t^{'},t^{'} - \frac{x^{'}}{v_F}) U(t - \frac{x}{v_F},t)\nonumber \\&\mbox{ }\mbox{ }\mbox{ }\mbox{ }\frac{1}{2}\bigg(< e^{ 2 \pi i \int^{x} (\rho_R(y,t) + \rho_L(-y,t) )  \mbox{      }dy } \mbox{ } e^{ -2 \pi i \int^{x^{'}} \rho_R(y^{'},t^{'}) dy^{'}}> \nonumber \\&\mbox{ }\mbox{ }\mbox{ }\mbox{ }+ < e^{ 2 \pi i \int^{x} \rho_R(y,t) \mbox{      }dy }\mbox{ } \mbox{ } e^{ -2 \pi i \int^{x^{'}} (\rho_R(y^{'},t^{'}) + \rho_L(-y^{'},t^{'}) )  \mbox{      }dy^{'} }> \bigg)
\label{rrhat}
\intertext{and for}
< T \mbox{ }\tilde{\psi}_R(x,t) \tilde{\psi}^{\dagger}_R(x^{'},t^{'})>_{\lambda=1}&= U(t^{'},t)\mbox{ }\frac{1}{2}\bigg(< e^{ 2 \pi i \int^{x} (\rho_R(y,t) + \rho_L(-y,t) )  \mbox{      }dy } \mbox{ } e^{ -2 \pi i \int^{x^{'}} \rho_R(y^{'},t^{'}) dy^{'}}> \nonumber \\&\mbox{ }\mbox{ }\mbox{ }\mbox{ }+ < e^{ 2 \pi i \int^{x} \rho_R(y,t) \mbox{      }dy }\mbox{ } \mbox{ } e^{ -2 \pi i \int^{x^{'}} (\rho_R(y^{'},t^{'}) + \rho_L(-y^{'},t^{'}) )  \mbox{      }dy^{'} }> \bigg)
\label{rrtilde}
\intertext{and we will also use the conventional choice in this case,}
< T \mbox{ }\tilde{\psi}_R(x,t) \tilde{\psi}^{\dagger}_R(x^{'},t^{'})>_{\lambda=0} &= U(t^{'},t)\mbox{ } <e^{ 2 \pi i  \int^{x} \rho_R (y,t) \mbox{      }dy }\mbox{ } e^{ -2 \pi i  \int^{x^{'}} \rho_{R}(y^{'},t^{'}) \mbox{      }dy^{'} } >
\intertext{The $LR$ correlations of the transformed fields are written as,}
< \tilde{\psi}_L(x,t)\hat{\psi}^{\dagger}_R(x^{'},t^{'})>_{\lambda=1}
&= \mbox{ }\frac{1}{2}\bigg(<  e^{ -2 \pi i \int^{x} \rho_L(y,t)\mbox{      } dy }\mbox{ } \mbox{ } e^{ -2 \pi i \int^{x^{'}} (\rho_R(y^{'},t^{'}) + \rho_L(-y^{'},t^{'}) )  \mbox{      }dy^{'} }>\mbox{ }+ \nonumber \\&\mbox{ }\mbox{ }\mbox{ }\mbox{ } <e^{ -2 \pi i \int^{x} ( \rho_L(y,t) + \rho_R(-y,t) ) \mbox{      } dy }\mbox{ }e^{ -2 \pi i \int^{x^{'}} \rho_R(y^{'},t^{'}) \mbox{      }dy^{'} }>\bigg)U(t^{'},t^{'}-\frac{x^{'}}{v_{F}})
\label{lrtildehat}
\intertext{and}
< \hat{\psi}_L(x,t)\tilde{\psi}^{\dagger}_R(x^{'},t^{'})>_{\lambda=1}
&= U(t^{'},t+\frac{x}{v_{F}})\mbox{ }\frac{1}{2}\bigg(<  e^{ -2 \pi i \int^{x} \rho_L(y,t)\mbox{      } dy }\mbox{ } \mbox{ } e^{ -2 \pi i \int^{x^{'}} (\rho_R(y^{'},t^{'}) + \rho_L(-y^{'},t^{'}) )  \mbox{      }dy^{'} }> \nonumber \\&\mbox{ }\mbox{ }\mbox{ }\mbox{ }+ <e^{ -2 \pi i \int^{x} ( \rho_L(y,t) + \rho_R(-y,t) ) \mbox{      } dy }\mbox{ }e^{ -2 \pi i \int^{x^{'}} \rho_R(y^{'},t^{'}) \mbox{      }dy^{'} }>\bigg)
\intertext{The choice of $\lambda=0$ is not valid in this case.}
\label{lrhattilde}
\intertext{The $RL$ correlations of the transformed fields can be written as}
<\hat{ \psi}_R(x,t) \tilde{\psi}^{\dagger}_L(x^{'},t^{'})>_{\lambda=1}
&= U(t-\frac{x}{v_{F}},t)\mbox{ }\frac{1}{2}\bigg(< e^{ 2 \pi i \int^{x} (\rho_R(y,t) + \rho_L(-y,t) )  \mbox{      }dy } \mbox{ }e^{ 2 \pi i \int^{x^{'}} \rho_L(y^{'},t^{'})\mbox{      } dy^{'} }> \nonumber \\& \mbox{ }\mbox{ }\mbox{ }\mbox{ }+ <  e^{ 2 \pi i \int^{x} \rho_R(y,t) \mbox{      }dy }\mbox{ } e^{ 2 \pi i \int^{x^{'}} ( \rho_L(y^{'},t^{'}) + \rho_R(-y^{'},t^{'}) ) \mbox{      } dy^{'} } > \bigg)
\label{rlhattilde}
\intertext{and}
<\tilde{ \psi}_R(x,t) \hat{\psi}^{\dagger}_L(x^{'},t^{'})>_{\lambda=1}
&= U(t^{'} + \frac{x^{'}}{v_{F}},t)\mbox{ }\frac{1}{2}\bigg(< e^{ 2 \pi i \int^{x} (\rho_R(y,t) + \rho_L(-y,t) )  \mbox{      }dy } \mbox{ }e^{ 2 \pi i \int^{x^{'}} \rho_L(y^{'},t^{'})\mbox{      } dy^{'} }> \nonumber \\& \mbox{ }\mbox{ }\mbox{ }\mbox{ }+ <  e^{ 2 \pi i \int^{x} \rho_R(y,t) \mbox{      }dy }\mbox{ } e^{ 2 \pi i \int^{x^{'}} ( \rho_L(y^{'},t^{'}) + \rho_R(-y^{'},t^{'}) ) \mbox{      } dy^{'} } > \bigg)
\label{rltildehat}
\intertext{Again in this case the choice of $\lambda=0$ is invalid.}
\intertext{The $LL$ correlations of the transformed fields are written as}
< \hat{\psi}_L(x,t) \hat{\psi}^{\dagger}_L(x^{'},t^{'})>_{\lambda=1}
&= U(t^{'} + \frac{x^{'}}{v_F},t+\frac{x}{v_{F}})\mbox{ }\frac{1}{2}\bigg(< e^{ -2 \pi i \int^{x} \rho_L(y,t)\mbox{      } dy }\mbox{ } e^{ 2 \pi i \int^{x^{'}} ( \rho_L(y^{'},t^{'}) + \rho_R(-y^{'},t^{'}) ) \mbox{      } dy^{'} }> \nonumber \\& \mbox{ }\mbox{ }\mbox{ }\mbox{ } + < e^{ -2 \pi i \int^{x} ( \rho_L(y,t) + \rho_R(-y,t) ) \mbox{      } dy }\mbox{ } e^{ 2 \pi i \int^{x^{'}} \rho_L(y^{'},t^{'})\mbox{      } dy^{'} }> \bigg)
\intertext{and}
< \tilde{\psi}_L(x,t) \tilde{\psi}^{\dagger}_L(x^{'},t^{'})>_{\lambda=1}
&= \frac{1}{2}\bigg(< e^{ -2 \pi i \int^{x} \rho_L(y,t)\mbox{      } dy }\mbox{ } e^{ 2 \pi i \int^{x^{'}} ( \rho_L(y^{'},t^{'}) + \rho_R(-y^{'},t^{'}) ) \mbox{      } dy^{'} }> \nonumber \\& \mbox{ }\mbox{ }\mbox{ }\mbox{ } + < e^{ -2 \pi i \int^{x} ( \rho_L(y,t) + \rho_R(-y,t) ) \mbox{      } dy }\mbox{ } e^{ 2 \pi i \int^{x^{'}} \rho_L(y^{'},t^{'})\mbox{      } dy^{'} }> \bigg)
\intertext{and we also use the conventional choice}
<T \tilde{\psi}_L(x,t) \tilde{\psi}^{\dagger}_L(x^{'},t^{'})>_{\lambda=0} &= <e^{- 2 \pi i  \int^{x} \rho_L (y,t) \mbox{      }dy }\mbox{ } e^{ 2 \pi i  \int^{x^{'}} \rho_{L}(y^{'},t^{'}) \mbox{      }dy^{'} } >
\end{align}

\subsection{RR Green function:}
The NCBT ansatz for the nonequilibrium $RR$ Green function is
\begin{align}
< T\mbox{ }\Psi_R(x,t) \Psi^{\dagger}_R(x^{'},t^{'})> \mbox{ }=\mbox{ }&\theta(-x)\theta(-x^{'})< T \mbox{ }\tilde{\psi}_R(x,t) \tilde{\psi}^{\dagger}_R(x^{'},t^{'})>_{\lambda=0} \nonumber \\&+\mbox{ } C_{1}< T\mbox{ } \hat{\psi}_R(x,t) \hat{\psi}^{\dagger}_R(x^{'},t^{'})>_{\lambda=1}+\mbox{ } C_{2}\mbox{ }<T\mbox{ } \tilde{\psi}_R(x,t) \tilde{\psi}^{\dagger}_R(x^{'},t^{'})>_{\lambda=1}
\label{RRfg}
\end{align}
Bosonization doesn't give us the prefactors so by comparing with the exact results we fix the constant prefactors (see \hyperref[AppendixA]{Appendix A} for details),
\begin{align}
C_{1} = \frac{i}{2 \beta v_{F}}\frac{\left(\frac{\Gamma}{v_{F}}\frac{  (2 v_{F})^2
}{\Gamma ^2 +4 v_F^2}\right)^{2}}{\sinh(\frac{\pi \epsilon}{\beta v_{F}}) \mbox{ }\mathscr{C}_{1}}\theta(x)\theta(x^{'})
\end{align}
where $e^{ \frac{1}{2} (2 \pi i)^2 \int^{x}  \mbox{      }dy\int^{x}  \mbox{      }dy^{'} \mbox{      } <(\rho_R(y,t) + \rho_L(-y,t) ) (\rho_R(y^{'},t) + \rho_L(-y^{'},t) ) > } = \sinh(\frac{\pi \epsilon}{\beta v_{F}})$ where $\epsilon$ is a regularization factor that is eventually taken to zero and we define $\mathscr{C}_{1} =  e^{ \frac{1}{2} (2 \pi i)^2  \int^{x^{'}}  \mbox{      }dy^{'} \int^{x^{'}}  \mbox{      }dy \mbox{      } < \rho_R(y^{'},t^{'})\rho_R(y,t^{'}) > }  $ with $x^{'} > 0$ which is ultimately independent of $x^{'}$ and $t^{'}$,
and
\begin{align}
C_{2} = \frac{i}{2 \beta v_{F}}\frac{\left( 1
-          \mbox{          }  \theta(x^{'}) \frac{  2 \Gamma^2
}{\Gamma ^2 +4 v_F^2}\right)\left( 1
-          \mbox{          }  \theta(x) \frac{  2 \Gamma^2
}{\Gamma ^2 +4 v_F^2}\right)-\theta(-x)\theta(-x^{'})}{\sinh(\frac{\pi \epsilon}{\beta v_{F}})\mbox{ }\frac{1}{2}((\theta(x)+\theta(x^{'}))\mathscr{C}_{1} + (\theta(-x)+\theta(-x^{'}))\sinh(\frac{\pi \epsilon}{\beta v_{F}})^{\frac{1}{2}} ) }
\end{align}
The term $\sinh(\frac{\pi \epsilon}{\beta v_{F}})^{\frac{1}{2}}$ comes from evaluating $e^{ \frac{1}{2} (2 \pi i)^2  \int^{x}  \mbox{      }dy^{'} \int^{x}  \mbox{      }dy \mbox{      } < \rho_R(y,t^{'})\rho_R(y^{'},t^{'}) > } $ with $x < 0$.
This form of $C_{1}$ and $C_{2}$ in Eq.\ref{RRfg} conspires to give us the correct constant prefactors as in the exact Green functions.\\\mbox{ }\\
\subsection{LR Green function:}
The $LR$ Green function  can be written as
\begin{align}
<T \mbox{ } \Psi_L(x,t)\Psi^{\dagger}_R(x^{'},t^{'})> \mbox{ }=\mbox{ } &C_{3}\mbox{ }<T \mbox{ } \tilde{\psi}_L(x,t)\hat{\psi}^{\dagger}_R(x^{'},t^{'})>_{\lambda=1} \mbox{ } -\mbox{ } C_{4} \mbox{ } <T \mbox{ } \hat{\psi}_L(x,t)\tilde{\psi}^{\dagger}_R(x^{'},t^{'})>_{\lambda=1}
\label{lrbose}
\end{align}
On comparison with the exact results we fix the constant prefactors to be (see \hyperref[AppendixB]{Appendix B} for details),
\begin{align}
C_{3} = -\frac{i}{2 \beta v_{F}}\frac{\left[ 1
-      \theta(-x )    \mbox{          }   \frac{  2 \Gamma^2
}{\Gamma ^2 +4 v_F^2}
\right]i \frac{\Gamma }{v_F}    \frac{ (  2  v_F )^2   }{\Gamma ^2 +4 v_F^2}}{\frac{1}{2}\sinh(\frac{\pi \epsilon}{\beta v_{F}})(\theta(-x) \mathscr{C}_{2} +\mathscr{C}_{1}+ \theta(x)\sinh(\frac{\pi \epsilon}{\beta v_{F}})^{\frac{1}{2}})}\theta(x^{'})
\end{align}
and
\begin{align}
C_{4} = -\frac{i}{2 \beta v_{F}}\frac{\left[1
-      \theta(x^{'})    \mbox{          }   \frac{  2 \Gamma^2
}{\Gamma ^2 +4 v_F^2}\right] i \frac{\Gamma }{v_F}    \frac{ (  2  v_F )^2   }{\Gamma ^2 +4 v_F^2}}{\frac{1}{2}\sinh(\frac{\pi \epsilon}{\beta v_{F}})(\theta(x^{'}) \mathscr{C}_{1} + \mathscr{C}_{2} + \theta(-x^{'})\sinh(\frac{\pi \epsilon}{\beta v_{F}})^{\frac{1}{2}})}\theta(-x)
\end{align}
where $\mathscr{C}_{2} = e^{ \frac{1}{2} (2 \pi i)^2  \int^{x^{'}}  \mbox{      }dy^{'} \int^{x^{'}}  \mbox{      }dy \mbox{      } < \rho_L(y,t^{'})\rho_L(y^{'},t^{'}) > }$ with $x^{'} < 0$ which is ultimately a constant independent of $x^{'}$ and $t^{'}$.
\subsection{RL Green function:}
The $RL$ Green function  can be written as
\begin{align}
<T \mbox{ } \Psi_R(x,t) \Psi^{\dagger}_L(x^{'},t^{'})>  \mbox{ }=\mbox{ } -C_{5}\mbox{ }<T \mbox{ }\hat{ \psi}_R(x,t) \tilde{\psi}^{\dagger}_L(x^{'},t^{'})>_{\lambda=1} \mbox{ }+ \mbox{ }C_{6}\mbox{ }<T \mbox{ }\tilde{ \psi}_R(x,t) \hat{\psi}^{\dagger}_L(x^{'},t^{'})>_{\lambda=1}
\label{rlbose}
\end{align}
Comparing with the exact results we can express,
\begin{align}
C_{5} = \frac{i}{2 \beta v_{F}}\frac{\left[ 1
-      \theta(-x^{'})    \mbox{          }   \frac{  2 \Gamma^2
}{\Gamma ^2 +4 v_F^2}
\right]i  \frac{\Gamma }{v_F}\frac{(2 v_F)^2
}{\Gamma ^2 +4 v_F^2}}{\frac{1}{2}\sinh(\frac{\pi \epsilon}{\beta v_{F}})(\theta(-x^{'}) \mathscr{C}_{2} + \mathscr{C}_{1}+ \theta(x^{'})\sinh(\frac{\pi \epsilon}{\beta v_{F}})^{\frac{1}{2}})}\theta(x)
\end{align}
and
\begin{align}
C_{6} = \frac{i}{2 \beta v_{F}}\frac{\left[1
-      \theta(x )    \mbox{          }   \frac{  2 \Gamma^2
}{\Gamma ^2 +4 v_F^2}\right]i \frac{\Gamma }{v_F}    \frac{ (  2  v_F )^2   }{\Gamma ^2 +4 v_F^2}}{\frac{1}{2}\sinh(\frac{\pi \epsilon}{\beta v_{F}})(\theta(x) \mathscr{C}_{1} +\mathscr{C}_{2}+ \theta(-x)\sinh(\frac{\pi \epsilon}{\beta v_{F}})^{\frac{1}{2}})} \theta(-x^{'})
\end{align}
\subsection{LL Green function:}
Finally the $LL$ Green function is expressed as,
\begin{align}
<T \mbox{ }\Psi_L(x,t) \Psi^{\dagger}_L(x^{'},t^{'})> \mbox{ }=\mbox{ }&\theta(x)\theta(x^{'})<T \tilde{\psi}_L(x,t) \tilde{\psi}^{\dagger}_L(x^{'},t^{'})>_{\lambda=0}\nonumber \\&+\mbox{ } C_{7}\mbox{ }<T\mbox{ } \hat{\psi}_L(x,t) \hat{\psi}^{\dagger}_L(x^{'},t^{'})>_{\lambda=1} \mbox{ }+ \mbox{ } C_{8}<T\mbox{ } \tilde{\psi}_L(x,t) \tilde{\psi}^{\dagger}_L(x^{'},t^{'})>_{\lambda=1}
\end{align}
Again on comparing with exact results we can express,
\begin{align}
C_{7} = -\frac{i}{2 \beta v_{F}}\frac{\left( \frac{\Gamma }{v_F} \mbox{          } \frac{(2 v_F)^2
}{\Gamma ^2 +4 v_F^2}\right)^2 }{\sinh(\frac{\pi \epsilon}{\beta v_{F}}) \mathscr{C}_{2}}\theta(-x)\theta(-x^{'})
\end{align}
and
\begin{align}
C_{8} = -\frac{i}{2 \beta v_{F}}\frac{\left[ 1
-      \theta(-x^{'})    \mbox{          }   \frac{  2 \Gamma^2
}{\Gamma ^2 +4 v_F^2}
\right]   \left[ 1
-      \theta(-x )    \mbox{          }   \frac{  2 \Gamma^2
}{\Gamma ^2 +4 v_F^2}
\right]-\theta(x)\theta(x^{'}) }{\frac{1}{2}\sinh(\frac{\pi \epsilon}{\beta v_{F}})((\theta(-x)+\theta(-x^{'})) \mathscr{C}_{2} + (\theta(x)+\theta(x^{'}))\sinh(\frac{\pi \epsilon}{\beta v_{F}})^{\frac{1}{2}})}
\end{align}
Although this bosonization procedure seems like a roundabout and complicated way to study an exactly solvable system (i.e. free fermions plus infinite bandwidth impurity and time dependent bias) it should prove to be very useful for future prospective research on systems with interparticle interactions like fractional quantum Hall edges out of equilibrium.
\section{FOUR-POINT FUNCTIONS}
\label{fourpoint}
In this section we evaluate four point-functions using our unconventional bosonization scheme. We still have not made clear why we have chosen only the conventional bosonization form (with $\lambda=0$) for the cases $<T \mbox{ } \Psi_R(x<0,t) \Psi^{\dagger}_R(x^{'}<0,t^{'})> $ and $<T \mbox{ } \Psi_L(x>0,t) \Psi^{\dagger}_L(x^{'}>0,t^{'})> $ whereas the unconventional choice ($\lambda=1$) superficially seems to work for all cases. Evaluating the four-point functions will provide clarity on this issue by means of some consistency checks. Since we are not considering interparticle interactions we can use Wick's theorem to write down the four-point functions. Let us first consider
\begin{align}
<T\mbox{ }\rho_R(x_1,t_1)\psi^{\dagger}_R(x,t)\psi_R(x^{'},t^{'})>\mbox{ }=\mbox{ }-<T\mbox{ }\psi_R(x^{'},t^{'})\psi^{\dagger}_R(x_1,t_{1})>  <T\mbox{ }\psi_R(x_1,t_1)\psi^{\dagger}_R(x,t)>
\label{rhoRfp}
\end{align}
We shall use the proposed bosonization scheme to reproduce the RHS of Eq.\ref{rhoRfp}. Expressing $\psi^{\dagger}_R(x,t)$ and $\psi_R(x^{'},t^{'})$ in bosonized form we can write this as
\begin{align}
-<T\mbox{ }&\psi_R(x^{'},t^{'})\psi^{\dagger}_R(x_1,t_{1})>  <T\mbox{ }\psi_R(x_1,t_1)\psi^{\dagger}_R(x,t)> \nonumber \\\mbox{         } = \mbox{          }&w_{0}\mbox{ }  <\rho_R(x_1,t_1) e^{ -2\pi i\int^{x} dy \mbox{ } \rho_R(y,t)  } e^{ 2\pi i\int^{x^{'}} dy^{'} \mbox{ } \rho_R(y^{'},t^{'}) } > \nonumber \\
&+ w_{11} \mbox{    } <\rho_R(x_1,t_1) e^{ -2\pi i\int^{x} dy \mbox{ } (\rho_R(y,t)+  \rho_L(-y,t)) } e^{ 2\pi i\int^{x^{'}} dy^{'} \mbox{ } \rho_R(y^{'},t^{'}) } >\nonumber \\
&+ w_{21} \mbox{    } <\rho_R(x_1,t_1) e^{ -2\pi i\int^{x} dy \mbox{ }  \rho_R(y,t) } e^{ 2\pi i\int^{x^{'}} dy^{'} \mbox{ } (\rho_R(y^{'},t^{'})+ \rho_L(-y^{'},t^{'})) } >
\label{Rfpbose}
\end{align}
where $w_{0}$ is the prefactor for the term with $\lambda=0$ in the exponent (conventional choice) and $w_{11}$ and $w_{12}$ are the prefactors for the terms with $\lambda=1$ in the exponent of $\psi^{\dagger}_R(x,t)$ and $\psi_R(x^{'},t^{'})$ respectively. Without loss of generality we can write $\rho_R(x_1,t_1) = \lim_{a \rightarrow 0}\frac{d}{da} e^{a\mbox{ }\rho_R(x_1,t_1)}$ and we consider only the most singular parts of the RHS of Eq.\ref{Rfpbose}. In order to reproduce the exact result for the four-point function obtained using Wick's theorem we come to the conclusion that the terms on the RHS with $\lambda=0$ are only valid for the case with $x<0, x^{'}<0$ (see \hyperref[AppendixC]{Appendix C} for details) while for all remaining cases only the unconventional bosonization terms with $\lambda=1$ will give us the correct results with the prefactors appropriately fixed as shown in \hyperref[AppendixC]{Appendix C}. For the case $x<0, x^{'}<0$ it appears at the outset that both the conventional as well as the unconventional bosonization choices should work, but evaluating the four-point function of the type $<T\mbox{ }\rho_L(-x_1,t_1)\psi^{\dagger}_R(x,t)\psi_R(x^{'},t^{'})> $ gives additional constraints that force the condition that the prefactors for the $\lambda=1$ terms be zero for the case $x<0, x^{'}<0$. Using Wick's theorem we write
\begin{align}
<T\mbox{ }\rho_L(-x_1,t_1)\psi^{\dagger}_R(x,t)\psi_R(x^{'},t^{'})>\mbox{ }=\mbox{ }-<T\mbox{ }\psi_R(x^{'},t^{'})\psi^{\dagger}_L(-x_1,t_1)>   <T\mbox{ }\psi_L(-x_1,t_{1})\psi^{\dagger}_R(x,t)>
\end{align}
Similar to Eq.\ref{Rfpbose} we can write this as,
\begin{align}
-<T\mbox{ }&\psi_R(x^{'},t^{'})\psi^{\dagger}_L(-x_1,t_1)>   <T\mbox{ }\psi_L(-x_1,t_{1})\psi^{\dagger}_R(x,t)> \nonumber \\ \mbox{ }=\mbox{ } &q_{0}\mbox{    } <\rho_L(-x_1,t_1) e^{ -2\pi i\int^{x} dy \mbox{ } \rho_R(y,t) } e^{ 2\pi i\int^{x^{'}} dy^{'} \mbox{ } \rho_R(y^{'},t^{'}) } > \nonumber \\
&+ q_{11}<\rho_L(-x_1,t_1) e^{ -2\pi i\int^{x} dy \mbox{ } (\rho_R(y,t)+\rho_L(-y,t)) } e^{ 2\pi i\int^{x^{'}} dy^{'} \mbox{ } \rho_R(y^{'},t^{'}) } > \nonumber \\
&+ q_{21} \mbox{    } <\rho_L(-x_1,t_1) e^{ -2\pi i\int^{x} dy \mbox{ }  \rho_R(y,t) } e^{ 2\pi i\int^{x^{'}} dy^{'} \mbox{ } (\rho_R(y^{'},t^{'})+\rho_L(-y^{'},t^{'})) } >
\end{align}
In \hyperref[AppendixC]{Appendix C} we show that in order to reproduce the correct Wick's theorem result for the case $x<0, x^{'}<0$, the prefactors $q_{11}$ and $q_{12}$ should be identically zero. This means that the unconventional bosonization terms with $\lambda = 1$ in the exponent do not play any role in this particular calculation. So we can argue on grounds of self consistency that when evaluating the two-point functions the conventional bosonization procedure is the only valid choice for the $< T\mbox{ }\Psi_R(x<0,t) \Psi^{\dagger}_R(x^{'}<0,t^{'})>$ correlations. A similar procedure evaluating the four-point functions $<T\mbox{ }\rho_R(-x_1,t_1)\psi^{\dagger}_L(x,t)\psi_L(x^{'},t^{'})>$ and $<T\mbox{ }\rho_L(x_1,t_1)\psi^{\dagger}_L(x,t)\psi_L(x^{'},t^{'})>$ can be used to show that for writing the bosonized form of $<T \Psi_L(x>0,t) \Psi^{\dagger}_L(x^{'}>0,t^{'})> $  the conventional bosonization formula is the only valid choice. But for all other cases it is necessary to use the unconventional bosonization scheme to obtain the correct form of the two-point correlations.
\section{TUNNELING CURRENT AND CONDUCTANCE}
\label{currentcond}
The tunneling current is defined in the usual sense as the rate of change of the difference in the number of right and left movers
\begin{align}
 I_{tun} = e\mbox{ }\partial_{t}\frac{\Delta N}{2} = e\frac{i}{2}\left[H,\Delta N\right] = e\frac{i}{2}\left[H,N_{R}-N_{L}\right]
 \label{tundef}
 \end{align}
From the Green functions obtained using our bosonization ansatz we obtain the expected results for the tunneling current \cite{Babu_2022,PhysRevB.54.7366,PhysRevB.84.155414},
 \begin{align}
  I_{tun} \mbox{        } &= \mbox{          }-i e \Gamma \mbox{        } \lim_{t^{'} \rightarrow t } \bigg( < \Psi^{\dagger}_R(0,t^{'})  \Psi_L(0,t) > - < \Psi^{\dagger}_L(0,t)  \Psi_R(0,t^{'}) > \bigg)
  \label{eqitun}
 \end{align}
 Using Eqs.\ref{rlbose} and \ref{lrbose} along with the bosonized correlations in Eqs.\ref{lrtildehat}-\ref{rltildehat} we get,
 \begin{align}
<\Psi^{\dagger}_{R}&(0,t^{'})\Psi_{L}(0,t)>   \mbox{        } = \mbox{          }
- \frac{i}{2\pi} \frac{ \frac{ \pi }{\beta v_F } }{\sinh( \frac{ \pi }{\beta v_F } (-v_F(t-t^{'}) ) )  }
 \left( - U(t^{'},t) + 1  \right)\mbox{ }i \frac{\Gamma }{v_F}    \frac{   2  v_F^2   }{\Gamma^2 +4 v_F^2} \left[1
-       \frac{    \Gamma^2
}{\Gamma ^2 +4 v_F^2}\right]
\end{align}
Note that we use the Dirichlet regularized step function ($\theta(0) = \frac{1}{2})$. This gives us
\begin{align}
 I_{tun} \mbox{        } &=  \mbox{          }i e \Gamma \mbox{        }  \lim_{t^{'} \rightarrow t }\mbox{        }  \bigg(   \frac{1}{2\pi} \frac{ \frac{ \pi }{\beta v_F } }{\sinh( \frac{ \pi }{\beta v_F } (v_F(t-t^{'}) ) )  }
 \left(  U(t,t^{'}) - U(t^{'},t)  \right) \mbox{          }   \frac{\Gamma }{v_F}    \frac{   2  v_F^2   }{\Gamma ^2 +4 v_F^2} \left[1
-       \frac{    \Gamma^2
}{\Gamma ^2 +4 v_F^2}\right] \bigg) \nonumber \\
&= \mbox{          }   - \Gamma^2 \mbox{        }
     \mbox{          }    \frac{   4   }{\Gamma ^2 +4 v_F^2} \left[1
-       \frac{    \Gamma^2
}{\Gamma ^2 +4 v_F^2}\right]
\mbox{    }
\frac{e^2}{2\pi } V_b
\end{align}
 We define the tunneling amplitude in terms of a tunneling parameter $t_{p}$ as $\Gamma = 2 v_{F} t_{p}$ \cite{PhysRevB.94.235426} and since we adopt the same convention as in our previous work we replace $V_{b} = -V$ and we get
 \begin{align}
 I_{tun} \mbox{        } = \mbox{          }
     \mbox{          }    \frac{   4 t_{p}^{2}  }{(t_{p}^{2} +1)^{2}}
\mbox{    }
\frac{e^2}{h } V
\label{eqtunc}
 \end{align}
 and the differential tunneling conductance is obtained as
 \begin{align}
 G = \frac{d I_{tun}}{d V} \mbox{ }=\mbox{ } G_{tun} \mbox{        } = \mbox{          }  \frac{   4 t_{p}^{2}  }{(t_{p}^{2} +1)^{2}}
\mbox{    }
\frac{e^2}{ h }
\label{eqcond}
 \end{align}
 These results agree with standard scattering theory \cite{LANDAUER198191,PhysRevB.31.6207} and also it is apparent that there is a duality between the strong and weak tunneling regimes ($t_{p} \rightarrow \frac{1}{t_{p}}$) in the tunneling conductance. The $I(V)$ characteristics is linear and nonlinearities in the transport are expected only when mutual interactions between the fermions are present or when a finite bandwidth point-contact is considered \cite{Babu_2022}.
\section{CONCLUSIONS AND PROSPECTS }
\label{conclusion}
In summary, we use a modified version of the Fermi-Bose correspondence (viz. Eq.\ref{ncbt}) and construct an ansatz to obtain the bosonized version of the nonequilibrium Green functions (NEGF) for two noninteracting chiral quantum wires coupled through a point contact at the origin with a bias between the right and left movers. The standard bosonization formalism cannot be employed easily as it does not yield the exact Green functions in a closed form for models with impurity backscattering with or without bias.  The purpose of this article is to lay the foundation for our future work which is to show that the non-chiral bosonization technique (NCBT) can be used to write down the most singular parts of the full NEGF for the above mentioned system (considered in \cite{PhysRevB.52.8934}) in terms of simple functions of position and time, when mutual interactions between fermions are included. Evaluating the full nonequilibrium Green function (NEGF) in presence of interparticle interactions is a highly nontrivial task. It is known that the edge states of fractional quantum Hall fluids can be described as chiral Luttinger liquids \cite{PhysRevB.41.12838,doi:10.1142/S0217979292000840,doi:10.1080/00018739500101566}. Due to the nature of the chiral Luttinger liquid a power-law dependence in the tunneling $I(V)$ characteristics is expected (at zero temperature). More generally, experimental evidence shows that the tunneling behaviour as a function of bias voltage $V$ or temperature $T$ follows a universal scaling form as predicted by theory. Fendley, Ludwig and Saleur \cite{PhysRevB.52.8934} have shown that the problem of tunneling between chiral Luttinger edges is integrable for certain values of the filling fractions (between $1/4$ and $1$) using thermodynamic Bethe ansatz and they obtain the universal scaling functions for the nonequilibrium tunneling current and conductance. We also show that the modified unconventional bosonization ansatz can be used to evaluate the relevant four-point functions consistent with Wick's theorem in the absence of interparticle interactions. This is a crucial cross-check that validates our method.\\

 In bosonization (both NCBT and conventional) inclusion of forward scattering is as easy (or difficult) as solving the theory without forward scattering. This is because bosonization is essentially describing fermions using commuting variables.  Ultimately forward scattering mutual interaction between the fermions results in just a Gaussian deformation of the theory (Hubbard Stratanovich transformation) without such interactions. But including the effects of impurity backscattering (the point-contact in this case) even without interparticle interactions is tremendously nontrivial. Being able to solve for the properties of free fermions but in presence of backward scattering from an impurity requires radically new approaches such as NCBT if one desires to treat the impurity properly. In our formalism, interparticle interactions can be treated non-perturbatively by calculating the density-density correlation functions (DDCF) using a generating function with an auxiliary field in a manner similar to the procedure in \cite{Danny_Babu_2020}. The interacting DDCF can be used in our bosonization ansatz for the nonequilibrium correlation functions. Once we have the interacting NEGF one can readily obtain the universal scaling form \footnote{There is an influential segment of the community that dismiss NCBT and refuse to believe it because the final general space-time Green functions derived using NCBT are non-universal (impurity strength dependent). They wrongly conclude that this means the scaling functions are also non-universal. We wish to show in the concluding part of this work that since the scaling functions are equal-space and equal-time limits of the full Green functions, the impurity strength dependence drops out making the scaling functions universal.} of the tunneling properties and this will hopefully prove to be a crucial validation for our technique. We expect our method to reproduce the universal scaling behaviour as the previously obtained NCBT Green's functions \cite{doi:10.1142/S0217751X18501749} for an interacting Luttinger liquid with impurities in equilibrium does indeed show power-law scaling behaviour in the equal space-equal time limit (see \hyperref[AppendixD]{Appendix D}). The central achievement of this paper is being able to describe fermions backward scattering from an impurity in presence of a bias that drives the system out of equilibrium exactly using commuting variables as opposed to anticommuting variables. In the former approach, inclusion of mutual interaction between fermions is a (relatively) trivial Gaussian deformation of these results (formidable practically). In the latter approach, inclusion of even forward scattering between fermions is intractable. Technically, being able to solve the free fermion theory using bosonization is the crux of the problem, even though nontrivial physical phenomena are seen upon inclusion of forward scattering between fermions. This crucial concluding part of the problem will be dealt with in a future work.


\section*{APPENDIX A: Calculation of the RR Green function from the unconventional bosonization ansatz}
\label{AppendixA}
\setcounter{equation}{0}
\renewcommand{\theequation}{A.\arabic{equation}}
In this appendix we show in detail the calculation of the $RR$ Green function from the unconventional bosonization (NCBT) ansatz. The $LL$ Green function is also obtained in a similar manner and hence is not shown separately. We have to begin with,
 \begin{align}
< T\mbox{ }\Psi_R(x,t) \Psi^{\dagger}_R(x^{'},t^{'})> \mbox{ }=\mbox{ }&\theta(-x)\theta(-x^{'})< T \mbox{ }\tilde{\psi}_R(x,t) \tilde{\psi}^{\dagger}_R(x^{'},t^{'})>_{\lambda=0} \nonumber \\&+\mbox{ } C_{1}< T\mbox{ } \hat{\psi}_R(x,t) \hat{\psi}^{\dagger}_R(x^{'},t^{'})>_{\lambda=1}+\mbox{ } C_{2}\mbox{ }<T\mbox{ } \tilde{\psi}_R(x,t) \tilde{\psi}^{\dagger}_R(x^{'},t^{'})>_{\lambda=1}
\end{align}
Using Eqs.\ref{rrhat} and \ref{rrtilde} we can write this as,
\begin{align}
< T\mbox{ }\Psi_R(x,t) \Psi^{\dagger}_R(x^{'},t^{'})> &=\theta(-x)\theta(-x^{'})U(t^{'},t)\mbox{ } <e^{ 2 \pi i  \int^{x} \rho_R (y,t) \mbox{      }dy }\mbox{ } e^{ -2 \pi i  \int^{x^{'}} \rho_{R}(y^{'},t^{'}) \mbox{      }dy^{'} } >\nonumber \\&\mbox{ }\mbox{ }\mbox{ }\mbox{ }+\mbox{ }C_{1}\mbox{ }U(t^{'},t^{'} - \frac{x^{'}}{v_F}) U(t - \frac{x}{v_F},t)\nonumber \\&\mbox{ }\mbox{ }\mbox{ }\mbox{ }\frac{1}{2}\bigg(< e^{ 2 \pi i \int^{x} (\rho_R(y,t) + \rho_L(-y,t) )  \mbox{      }dy } \mbox{ } e^{ -2 \pi i \int^{x^{'}} \rho_R(y^{'},t^{'}) dy^{'}}> \nonumber \\&\mbox{ }\mbox{ }\mbox{ }\mbox{ }+ < e^{ 2 \pi i \int^{x} \rho_R(y,t) \mbox{      }dy }\mbox{ } \mbox{ } e^{ -2 \pi i \int^{x^{'}} (\rho_R(y^{'},t^{'}) + \rho_L(-y^{'},t^{'}) )  \mbox{      }dy^{'} }> \bigg)
\nonumber \\&\mbox{ }\mbox{ }\mbox{ }+\mbox{ }C_{2}\mbox{ }U(t^{'},t)\mbox{ }\frac{1}{2}\bigg(< e^{ 2 \pi i \int^{x} (\rho_R(y,t) + \rho_L(-y,t) )  \mbox{      }dy } \mbox{ } e^{ -2 \pi i \int^{x^{'}} \rho_R(y^{'},t^{'}) dy^{'}}> \nonumber \\&\mbox{ }\mbox{ }\mbox{ }\mbox{ }+ < e^{ 2 \pi i \int^{x} \rho_R(y,t) \mbox{      }dy }\mbox{ } \mbox{ } e^{ -2 \pi i \int^{x^{'}} (\rho_R(y^{'},t^{'}) + \rho_L(-y^{'},t^{'}) )  \mbox{      }dy^{'} }> \bigg)
\label{rrappendix}
\end{align}
Let us now evaluate the expectation using a version of the Baker-Campbell-Haussdorff formula (Eq. \ref{bch}),
\begin{align}
< &e^{ 2 \pi i \int^{x} (\rho_R(y,t) + \rho_L(-y,t) )  \mbox{      }dy } \mbox{ } e^{ -2 \pi i \int^{x^{'}} \rho_R(y^{'},t^{'}) dy^{'}}> \nonumber \\ &= e^{\frac{1}{2}(2 \pi i)^{2}\int^{x}dy \int^{x} dy^{'} <(\rho_R(y,t) + \rho_L(-y,t) ) (\rho_R(y^{'},t) + \rho_L(-y^{'},t) )> }\nonumber \\
&\mbox{ }\mbox{ }\mbox{ }\mbox{ }e^{\frac{1}{2}(2 \pi i)^{2} \int^{x^{'}} dy \int^{x^{'}} dy^{'} <\rho_R(y,t^{'})\rho_R(y^{'},t^{'})>}e^{-(2 \pi i)^{2} \int^{x} dy \int^{x^{'}} dy^{'} <(\rho_R(y,t) + \rho_L(-y,t) )\rho_R(y^{'},t^{'})>}
\end{align}
Using the form of the density-density correlation functions in Sec.\ref{secdensity} we evaluate the integrals in the exponents,
\begin{align}
< &e^{ 2 \pi i \int^{x} (\rho_R(y,t) + \rho_L(-y,t) )  \mbox{      }dy } \mbox{ } e^{ -2 \pi i \int^{x^{'}} \rho_R(y^{'},t^{'}) dy^{'}}> \nonumber \\ &= \sinh(\frac{\pi \epsilon}{\beta v_{F}}) \mbox{ }(\theta(x^{'})\mathscr{C}_{1} + \theta(-x^{'})\sinh(\frac{\pi \epsilon}{\beta v_{F}})^{\frac{1}{2}})\mbox{ }\frac{1}{\sinh( \frac{ \pi }{\beta v_F } ( x-x^{'}-v_F(t-t^{'}) ) )}
\label{c11}
\end{align}
Here $\epsilon$ is a regularization factor that will eventually be taken to zero ($\epsilon \rightarrow 0$) and $\mathscr{C}_{1} =  e^{ \frac{1}{2} (2 \pi i)^2  \int^{x^{'}}  \mbox{      }dy^{'} \int^{x^{'}}  \mbox{      }dy \mbox{      } < \rho_R(y^{'},t^{'})\rho_R(y,t^{'}) > }  $ evaluated with $x^{'} > 0$  is ultimately a constant term. Similarly we get,
\begin{align}
< &e^{ 2 \pi i \int^{x} \rho_R(y,t) \mbox{      }dy }\mbox{ } \mbox{ } e^{ -2 \pi i \int^{x^{'}} (\rho_R(y^{'},t^{'}) + \rho_L(-y^{'},t^{'}) )  \mbox{      }dy^{'} }> \nonumber \\ &= \sinh(\frac{\pi \epsilon}{\beta v_{F}}) \mbox{ }(\theta(x)\mathscr{C}_{1} + \theta(-x)\sinh(\frac{\pi \epsilon}{\beta v_{F}})^{\frac{1}{2}})\mbox{ }\frac{1}{\sinh( \frac{ \pi }{\beta v_F } ( x-x^{'}-v_F(t-t^{'}) ) )}
\label{c12}
\end{align}
On comparing with the exact Green functions in Eq.\ref{eqnoneq} we see that for the first term the constant prefactors are of the form $\frac{i}{2 \beta v_{F}}\left(\frac{\Gamma}{v_{F}}\frac{  (2 v_{F})^2
}{\Gamma ^2 +4 v_F^2}\right)^{2}\theta(x)\theta(x^{'})$ and since bosonization doesn't give the prefactors we fix $C_{1}$ such that
\begin{align}
C_{1} = \frac{i}{2 \beta v_{F}}\frac{\left(\frac{\Gamma}{v_{F}}\frac{  (2 v_{F})^2
}{\Gamma ^2 +4 v_F^2}\right)^{2}}{\sinh(\frac{\pi \epsilon}{\beta v_{F}}) \mbox{ }\mathscr{C}_{1}}\theta(x)\theta(x^{'})
\end{align}
so that the term in the denominator cancels with a corresponding term that comes from evaluating the expectation values of the bosonized fields (Eqs.\ref{c11} and \ref{c12}) and substituting in Eq.\ref{rrappendix}. Note that the $\sinh(\frac{\pi \epsilon}{\beta v_{F}})^{\frac{1}{2}}$ term drops out since $C_{1}$ is non-zero only for $x,x^{'} > 0$.\\
Now in the second term the prefactor is $\frac{i}{2 \beta v_{F}}\left( 1
-          \mbox{          }  \theta(x^{'}) \frac{  2 \Gamma^2
}{\Gamma ^2 +4 v_F^2}\right)\left( 1
-          \mbox{          }  \theta(x) \frac{  2 \Gamma^2
}{\Gamma ^2 +4 v_F^2}\right)$ , hence we fix $C_{2}$ to be
\begin{align}
C_{2} = \frac{i}{2 \beta v_{F}}\frac{\left( 1
-          \mbox{          }  \theta(x^{'}) \frac{  2 \Gamma^2
}{\Gamma ^2 +4 v_F^2}\right)\left( 1
-          \mbox{          }  \theta(x) \frac{  2 \Gamma^2
}{\Gamma ^2 +4 v_F^2}\right)-\theta(-x)\theta(-x^{'})}{\sinh(\frac{\pi \epsilon}{\beta v_{F}})\mbox{ }\frac{1}{2}((\theta(x)+\theta(x^{'}))\mathscr{C}_{1} + (\theta(-x)+\theta(-x^{'}))\sinh(\frac{\pi \epsilon}{\beta v_{F}})^{\frac{1}{2}} ) }
\end{align}
The term in the denominator cancels with the same term appears from substituting Eqs.\ref{c11} and \ref{c12} in Eq.\ref{rrappendix}. So evaluating Eq.\ref{rrappendix} reproduces the correct exact $RR$ Green function as in Eq.\ref{eqnoneq}.
\begin{align}
<T\mbox{ } \Psi_R(x,t) \Psi^{\dagger}_R(x^{'},t^{'})> &= \bigg(\frac{i}{2 \beta v_{F}}\left(\frac{\Gamma}{v_{F}}\frac{  (2 v_{F})^2
}{\Gamma ^2 +4 v_F^2}\right)^{2}\theta(x)\theta(x^{'})\mbox{ }U(t^{'},t^{'} - \frac{x^{'}}{v_F}) U(t - \frac{x}{v_F},t) \nonumber \\&
\mbox{ }\mbox{ }\mbox{ }\mbox{ }+ \frac{i}{2 \beta v_{F}}\left( 1
-          \mbox{          }  \theta(x^{'}) \frac{  2 \Gamma^2
}{\Gamma ^2 +4 v_F^2}\right)\left( 1
-          \mbox{          }  \theta(x) \frac{  2 \Gamma^2
}{\Gamma ^2 +4 v_F^2}\right)\mbox{ }U(t^{'},t)\bigg)\nonumber \\&
\mbox{ }\mbox{ }\mbox{ }\mbox{ }csch( \frac{ \pi }{\beta v_F } ( x-x^{'}-v_F(t-t^{'}) ) )
\end{align}
A similar calculation gives us the correct $LL$ Green function as well.
\section*{APPENDIX B: Calculation of the LR Green function from the unconventional bosonization ansatz}
\label{AppendixB}
\setcounter{equation}{0}
\renewcommand{\theequation}{B.\arabic{equation}}
Here we show the calculation of the $LR$ Green function from the unconventional bosonization ansatz. The $RL$ Green function is also obtained in a similar manner and hence is not shown separately. We have
\begin{align}
<T \mbox{ } \Psi_L(x,t)\Psi^{\dagger}_R(x^{'},t^{'})> \mbox{ }=\mbox{ } &C_{3}\mbox{ }<T \mbox{ } \tilde{\psi}_L(x,t)\hat{\psi}^{\dagger}_R(x^{'},t^{'})>_{\lambda=1} \mbox{ } -\mbox{ } C_{4} \mbox{ } <T \mbox{ } \hat{\psi}_L(x,t)\tilde{\psi}^{\dagger}_R(x^{'},t^{'})>_{\lambda=1}
\end{align}
Using Eqns.\ref{lrtildehat} and \ref{lrhattilde} we write,
\begin{align}
<T \mbox{ } \Psi_L(x,t)\Psi^{\dagger}_R(x^{'},t^{'})> &= C_{3} \mbox{ }U(t^{'},t^{'}-\frac{x^{'}}{v_{F}})\mbox{ }\frac{1}{2}\bigg(<  e^{ -2 \pi i \int^{x} \rho_L(y,t)\mbox{      } dy }\mbox{ } \mbox{ } e^{ -2 \pi i \int^{x^{'}} (\rho_R(y^{'},t^{'}) + \rho_L(-y^{'},t^{'}) )  \mbox{      }dy^{'} }> \nonumber \\&\mbox{ }\mbox{ }\mbox{ }\mbox{ }+ <e^{ -2 \pi i \int^{x} ( \rho_L(y,t) + \rho_R(-y,t) ) \mbox{      } dy }\mbox{ }e^{ -2 \pi i \int^{x^{'}} \rho_R(y^{'},t^{'}) \mbox{      }dy^{'} }>\bigg)\nonumber \\
&\mbox{ }\mbox{ }\mbox{ }\mbox{ }- C_{4}\mbox{ } U(t^{'},t+\frac{x}{v_{F}})\mbox{ }\frac{1}{2}\bigg(<  e^{ -2 \pi i \int^{x} \rho_L(y,t)\mbox{      } dy }\mbox{ } \mbox{ } e^{ -2 \pi i \int^{x^{'}} (\rho_R(y^{'},t^{'}) + \rho_L(-y^{'},t^{'}) )  \mbox{      }dy^{'} }> \nonumber \\&\mbox{ }\mbox{ }\mbox{ }\mbox{ }+ <e^{ -2 \pi i \int^{x} ( \rho_L(y,t) + \rho_R(-y,t) ) \mbox{      } dy }\mbox{ }e^{ -2 \pi i \int^{x^{'}} \rho_R(y^{'},t^{'}) \mbox{      }dy^{'} }>\bigg)
\label{lrappendix}
\end{align}
Let us evaluate
\begin{align}
< & e^{ -2 \pi i \int^{x} \rho_L(y,t)\mbox{      } dy }\mbox{ } \mbox{ } e^{ -2 \pi i \int^{x^{'}} (\rho_R(y^{'},t^{'}) + \rho_L(-y^{'},t^{'}) )  \mbox{      }dy^{'} }> \nonumber \\
=&\mbox{ }\mbox{ } e^{\frac{1}{2}(2 \pi i)^{2} \int^{x}dy \int^{x} dy^{'} <\rho_{L}(y,t)\rho_{L}(y^{'},t)>}e^{\frac{1}{2}(2 \pi i)^{2} \int^{x^{'}}dy \int^{x^{'}} dy^{'} <(\rho_R(y,t^{'}) + \rho_L(-y,t^{'}) )(\rho_R(y^{'},t^{'}) + \rho_L(-y^{'},t^{'}) )>} \nonumber \\
&\mbox{ }\mbox{ }e^{(2 \pi i)^{2} \int^{x} dy \int^{x^{'}} dy^{'} <\rho_L(y,t)(\rho_R(y^{'},t^{'}) + \rho_L(-y^{'},t^{'}) )>}
\end{align}
Using the form of the density-density correlation functions in Sec.\ref{secdensity} we evaluate the integrals in the exponents,
\begin{align}
< & e^{ -2 \pi i \int^{x} \rho_L(y,t)\mbox{      } dy }\mbox{ } \mbox{ } e^{ -2 \pi i \int^{x^{'}} (\rho_R(y^{'},t^{'}) + \rho_L(-y^{'},t^{'}) )  \mbox{      }dy^{'} }> \nonumber \\
&= \sinh(\frac{\pi \epsilon}{\beta v_{F}})(\theta(-x) \mathscr{C}_{2} + \theta(x)\sinh(\frac{\pi \epsilon}{\beta v_{F}})^{\frac{1}{2}})\frac{1}{\sinh(\frac{ \pi }{\beta v_F } ( x+x^{'}+v_F(t-t^{'}) ))}
\label{c31}
\end{align}
Here $\epsilon$ is a regularization factor that is eventually made zero and $\mathscr{C}_{2} = e^{ \frac{1}{2} (2 \pi i)^2  \int^{x^{'}}  \mbox{      }dy^{'} \int^{x^{'}}  \mbox{      }dy \mbox{      } < \rho_L(y,t^{'})\rho_L(y^{'},t^{'}) > }$ with $x^{'}<0$ is a constant term. Similarly we get
\begin{align}
<&e^{ -2 \pi i \int^{x} ( \rho_L(y,t) + \rho_R(-y,t) ) \mbox{      } dy }\mbox{ }e^{ -2 \pi i \int^{x^{'}} \rho_R(y^{'},t^{'}) \mbox{      }dy^{'} }> \nonumber \\
&= \sinh(\frac{\pi \epsilon}{\beta v_{F}})(\theta(x^{'}) \mathscr{C}_{1} + \theta(-x^{'})\sinh(\frac{\pi \epsilon}{\beta v_{F}})^{\frac{1}{2}})\frac{1}{\sinh(\frac{ \pi }{\beta v_F } ( x+x^{'}+v_F(t-t^{'}) ))}
\label{c32}
\end{align}
On comparing with the exact Green functions in Eq.\ref{eqnoneq} we fix
\begin{align}
C_{3} = -\frac{i}{2 \beta v_{F}}\frac{\left[ 1
-      \theta(-x )    \mbox{          }   \frac{  2 \Gamma^2
}{\Gamma ^2 +4 v_F^2}
\right]i \frac{\Gamma }{v_F}    \frac{ (  2  v_F )^2   }{\Gamma ^2 +4 v_F^2}}{\frac{1}{2}\sinh(\frac{\pi \epsilon}{\beta v_{F}})(\theta(-x) \mathscr{C}_{2} +\mathscr{C}_{1}+ \theta(x)\sinh(\frac{\pi \epsilon}{\beta v_{F}})^{\frac{1}{2}})}\theta(x^{'})
\end{align}
such that the terms in the denominator cancel with corresponding terms that appear from substituting Eqs.\ref{c31} and \ref{c32} in Eq.\ref{lrappendix} and we get the correct prefactors. For the other prefactor we fix it to be
\begin{align}
C_{4} = -\frac{i}{2 \beta v_{F}}\frac{\left[1
-      \theta(x^{'})    \mbox{          }   \frac{  2 \Gamma^2
}{\Gamma ^2 +4 v_F^2}\right] i \frac{\Gamma }{v_F}    \frac{ (  2  v_F )^2   }{\Gamma ^2 +4 v_F^2}}{\frac{1}{2}\sinh(\frac{\pi \epsilon}{\beta v_{F}})(\theta(x^{'}) \mathscr{C}_{1} + \mathscr{C}_{2} + \theta(-x^{'})\sinh(\frac{\pi \epsilon}{\beta v_{F}})^{\frac{1}{2}})}\theta(-x)
\end{align}
so that the terms in the denominator cancel with corresponding terms that appear from substituting Eqs.\ref{c31} and \ref{c32} in Eq.\ref{lrappendix} and we get the correct prefactors. So upon evaluating Eq.\ref{lrappendix} we obtain the exact $LR$ Green functions as in Eq.\ref{eqnoneq},
\begin{align}
<T\mbox{ } \Psi_L(x,t)\Psi^{\dagger}_R(x^{'},t^{'})> &= \bigg(\frac{i}{2 \beta v_{F}}\left[ 1
-      \theta(-x )    \mbox{          }   \frac{  2 \Gamma^2
}{\Gamma ^2 +4 v_F^2}
\right]i \frac{\Gamma }{v_F}    \frac{ (  2  v_F )^2   }{\Gamma ^2 +4 v_F^2}\theta(x^{'})\mbox{ }U(t^{'},t^{'}-\frac{x^{'}}{v_{F}}) \nonumber \\
&\mbox{ }\mbox{ }\mbox{ }\mbox{ }- \frac{i}{2 \beta v_{F}}\left[1
-      \theta(x^{'})    \mbox{          }   \frac{  2 \Gamma^2
}{\Gamma ^2 +4 v_F^2}\right] i \frac{\Gamma }{v_F}    \frac{ (  2  v_F )^2   }{\Gamma ^2 +4 v_F^2}\theta(-x)\mbox{ }U(t^{'},t+\frac{x}{v_{F}})\bigg) \nonumber \\
&\mbox{ }\mbox{ }\mbox{ }\mbox{ }csch( \frac{ \pi }{\beta v_F } ( -x-x^{'}-v_F(t-t^{'}) ) )
\end{align}
\section*{APPENDIX C: Evaluating the four-point functions using unconventional bosonization ansatz}
\label{AppendixC}
\setcounter{equation}{0}
\renewcommand{\theequation}{C.\arabic{equation}}
First let us consider the general form of following four-point function,
\begin{align}
<T\mbox{ }&\rho_R(x_1,t_1)\psi^{\dagger}_R(x,t)\psi_R(x^{'},t^{'})>  \nonumber \\ \mbox{         } = \mbox{          }&
-<T\mbox{ }\psi_R(x^{'},t^{'})\psi^{\dagger}_R(x_1,t_{1+})>  <T\mbox{ }\psi_R(x_1,t_1)\psi^{\dagger}_R(x,t)> \nonumber \\\mbox{         } = \mbox{          }&
w_{0} \mbox{    } <\rho_R(x_1,t_1) e^{ -2\pi i\int^{x} dy \mbox{ } \rho_R(y,t) } e^{ 2\pi i\int^{x^{'}} dy^{'} \mbox{ } \rho_R(y^{'},t^{'}) } > \nonumber \\
&+ w_{11} \mbox{    } <\rho_R(x_1,t_1) e^{ -2\pi i\int^{x} dy \mbox{ } (\rho_R(y,t)+ \rho_L(-y,t)) } e^{ 2\pi i\int^{x^{'}} dy^{'} \mbox{ } \rho_R(y^{'},t^{'}) } >\nonumber \\
&+ w_{21} \mbox{    } <\rho_R(x_1,t_1) e^{ -2\pi i\int^{x} dy \mbox{ }  \rho_R(y,t) } e^{ 2\pi i\int^{x^{'}} dy^{'} \mbox{ } (\rho_R(y^{'},t^{'})+ \rho_L(-y^{'},t^{'})) } >
\end{align}
where $w_{0}$, $w_{11}$ and $w_{21}$ are prefactors.
We can write $\rho_R(x_1,t_1) = \lim_{a \rightarrow 0}\frac{d}{da} e^{a\mbox{ }\rho_R(x_1,t_1)}$ hence we get,
\begin{align}
-<T\mbox{ }&\psi_R(x^{'},t^{'})\psi^{\dagger}_R(x_1,t_{1})>  <T\mbox{ }\psi_R(x_1,t_1)\psi^{\dagger}_R(x,t)> \nonumber \\\mbox{ }=\mbox{ }
& w_{0} \mbox{    } \substack{ ( -2\pi i\int^{x} dy \mbox{ } <T\mbox{ }\rho_R(x_1,t_1)\rho_R(y,t)> + 2\pi i\int^{x^{'}} dy^{'} \mbox{ }<T\mbox{ }\rho_R(x_1,t_1) \rho_R(y^{'},t^{'})> )}\mbox{ } \mbox{  }\nonumber \\&
\mbox{  } e^{
\frac{1}{2} (2\pi i)^2 \int^{x} dy \mbox{ } \int^{x} dy^{'} \mbox{ }  <T\mbox{ }\rho_R(y,t)\rho_R(y^{'},t) >  } e^{
\frac{1}{2}( 2\pi i)^2 \int^{x^{'}} dy^{'} \mbox{ }\int^{x^{'}} dy  \mbox{ } <T\mbox{ }\rho_R(y^{'},t^{'})\rho_R(y,t^{'})> }\nonumber \\&
e^{ -(2\pi i)^2\int^{x} dy \mbox{ }    \int^{x^{'}} dy^{'} \mbox{ }<T\rho_R(y,t) \rho_R(y^{'},t^{'})> }\nonumber \\
&+  w_{11} \mbox{    } \substack{( -2\pi i\int^{x} dy \mbox{ } <T\mbox{ }\rho_R(x_1,t_1) (\rho_R(y,t)+  \rho_L(-y,t))> + 2\pi i\int^{x^{'}} dy^{'} \mbox{ }<T \mbox{ }\rho_R(x_1,t_1)  \rho_R(y^{'},t^{'})> )}
\nonumber \\&
\mbox{    }  e^{ \frac{1}{2}(2\pi i)^2 \int^{x} dy \mbox{ }\int^{x} dy^{'} \mbox{ } <T \mbox{ }(\rho_R(y,t)+  \rho_L(-y,t)) (\rho_R(y^{'},t)+  \rho_L(-y^{'},t))> } e^{ \frac{1}{2} (2\pi i)^2 \int^{x^{'}} dy^{'} \mbox{ } \int^{x^{'}} dy \mbox{ }<T \mbox{ }\rho_R(y^{'},t^{'}) \rho_R(y,t^{'})> }\nonumber \\&  \mbox{    }  e^{ -(2\pi i)^2 \int^{x} dy \mbox{ }  \int^{x^{'}} dy^{'} \mbox{ } <T \mbox{ }(\rho_R(y,t)+  \rho_L(-y,t))  \rho_R(y^{'},t^{'})> }\nonumber \\
& + w_{21} \mbox{    }\substack{( -2\pi i\int^{x} dy \mbox{ }   <T \mbox{ }\rho_R(x_1,t_1) \rho_R(y,t)>  +  2\pi i\int^{x^{'}} dy^{'} \mbox{ } <T \mbox{ }\rho_R(x_1,t_1) (\rho_R(y^{'},t^{'})+ \rho_L(-y^{'},t^{'}))> )}\nonumber \\&
\mbox{  }e^{ \frac{1}{2}(2\pi i)^2 \int^{x} dy \mbox{ }\int^{x} dy^{'}
 \mbox{ }   <T \mbox{ }\rho_R(y,t) \rho_R(y^{'},t)> } e^{ \frac{1}{2} ( 2\pi i)^2 \int^{x^{'}} dy^{'} \mbox{ } \int^{x^{'}} dy\mbox{ }  <T \mbox{ }(\rho_R(y^{'},t^{'})+ \rho_L(-y^{'},t^{'}))(\rho_R(y,t^{'})+ \rho_L(-y,t^{'}))> }\nonumber \\& \mbox{    }    e^{ -(2\pi i)^2 \int^{x} dy \mbox{ }   \int^{x^{'}} dy^{'} \mbox{ }<T \mbox{ }\rho_R(y,t) (\rho_R(y^{'},t^{'})+ \rho_L(-y^{'},t^{'}))> }
 \label{rhoRfpc1}
\end{align}
We define the symmetric and antisymmetric density fields as follows
\begin{align}
\rho_{sym}(x,t) \equiv \rho_{R}(x,t)+\rho_{L}(-x,t)\nonumber \\
\rho_{asy}(x,t) \equiv \rho_{R}(x,t)-\rho_{L}(-x,t)
\end{align}
This means that we can write
\begin{align}
\rho_{R}(x,t) = \frac{\rho_{sym}(x,t)+\rho_{asy}(x,t) }{2} \nonumber \\
\rho_{L}(-x,t) = \frac{\rho_{sym}(x,t)-\rho_{asy}(x,t) }{2}
\end{align}
Using this in Eq.\ref{rhoRfpc1} we get,
\begin{align}
-<T\mbox{ }&\psi_R(x^{'},t^{'})\psi^{\dagger}_R(x_1,t_{1})>  <T\mbox{ }\psi_R(x_1,t_1)\psi^{\dagger}_R(x,t)> \nonumber \\\mbox{ }=\mbox{ }
& w_{0} \mbox{    } \substack{ ( -2\pi i\int^{x} dy \mbox{ }\frac{1}{4}
(<T \rho_{sym}(x_1,t_1)\rho_{sym}(y,t)>+<T\rho_{asy}(x_1,t_1)\rho_{asy}(y,t)> )}
 \nonumber \\&\substack{+ 2\pi i\int^{x^{'}} dy^{'} \mbox{ }\frac{1}{4}
 (<T\rho_{sym}(x_1,t_1)\rho_{sym}(y^{'},t^{'})>+<T\rho_{asy}(x_1,t_1)\rho_{asy}(y^{'},t^{'})> ))}\mbox{ }\nonumber \\
 &\mbox{  } e^{
\frac{1}{2} (2\pi i)^2 \frac{1}{4}\int^{x} dy \mbox{ } \int^{x} dy^{'} \mbox{ }(<T\rho_{sym}(y,t)\rho_{sym}(y^{'})>+<T\rho_{asy}(y,t)\rho_{asy}(y^{'},t)>)
  }\nonumber \\
  & e^{
\frac{1}{2}( 2\pi i)^2 \frac{1}{4} \int^{x^{'}} dy^{'} \mbox{ }\int^{x^{'}} dy  \mbox{ }
(<T\rho_{sym}(y^{'},t^{'})\rho_{sym}(y,t^{'})>+<T\rho_{asy}(y^{'},t^{'}) \rho_{asy}(y,t^{'})> )
}\nonumber \\
&e^{ -(2\pi i)^2 \frac{1}{4}\int^{x} dy \mbox{ }    \int^{x^{'}} dy^{'} \mbox{ }
(<T\rho_{sym}(y,t)\rho_{sym}(y^{'},t^{'})>+<T\rho_{asy}(y,t)\rho_{asy}(y^{'},t^{'})> )
}\nonumber \\
&+ w_{11} \mbox{    } \substack{( -2\pi i\int^{x} dy \mbox{ }
\frac{1}{2}< T\rho_{sym}(x_1,t_1) \rho_{sym}(y,t) >}
 \nonumber \\&\substack{+ 2\pi i \frac{1}{4}\int^{x^{'}} dy^{'} \mbox{ }
 ( <T\rho_{sym}(x_1,t_1)\rho_{sym}(y^{'},t^{'})>+<T\rho_{asy}(x_1,t_1)\rho_{asy}(y^{'},t^{'})>) )}\nonumber \\
  &\mbox{    }  e^{ \frac{1}{2}(2\pi i)^2 \int^{x} dy \mbox{ }\int^{x} dy^{'} \mbox{ }
  <T\rho_{sym}(y,t) \rho_{sym}(y^{'},t)> } \nonumber \\&e^{ \frac{1}{2} (2\pi i)^2 \frac{1}{4}
  \int^{x^{'}} dy^{'} \mbox{ } \int^{x^{'}} dy \mbox{ }
   ( < T\rho_{sym}(y^{'},t^{'})\rho_{sym}(y,t^{'}) >  + <T \rho_{asy}(y^{'},t^{'})\rho_{asy}(y,t^{'}) > )
  }\nonumber \\
  &\mbox{    }  e^{ -(2\pi i)^2 \frac{1}{2}\int^{x} dy \mbox{ }  \int^{x^{'}} dy^{'} \mbox{ } <T\rho_{sym}(y,t)  \rho_{sym}(y^{'},t^{'})>  }\nonumber \\
  &+ w_{21} \mbox{    }\substack{( -2\pi i \frac{1}{4}\int^{x} dy \mbox{ }   (
<T \rho_{sym}(x_1,t_1)\rho_{sym}(y,t) >+ <T \rho_{asy}(x_1,t_1)\rho_{asy}(y,t)>  )}
\nonumber \\&\substack{+  2\pi i \frac{1}{2}\int^{x^{'}} dy^{'} \mbox{ }  <T \rho_{sym}(x_1,t_1)\rho_{sym}(y^{'},t^{'})>}
  )\nonumber \\
  &e^{ \frac{1}{2}(2\pi i)^2 \frac{1}{4}\int^{x} dy \mbox{ }\int^{x} dy^{'}
 \mbox{ }   ( < T \rho_{sym}(y,t)\rho_{sym}(y^{'},t) > + <T \rho_{asy}(y,t)\rho_{asy}(y^{'},t)>)  } \nonumber \\&e^{ \frac{1}{2} ( 2\pi i)^2 \int^{x^{'}} dy^{'} \mbox{ } \int^{x^{'}} dy\mbox{ }  <T\rho_{sym}(y^{'},t^{'})\rho_{sym}(y,t^{'})> }\nonumber \\
 &\mbox{    }    e^{ -(2\pi i)^2 \frac{1}{2} \int^{x} dy \mbox{ }   \int^{x^{'}} dy^{'} \mbox{ }  <T\rho_{sym}(y,t) \rho_{sym}(y^{'},t^{'})>  }
\label{rhoRC5}
\end{align}
The correlations of the symmetric and antisymmetric fields are,
\begin{align}
<T \mbox{ }\rho_{sym}(x,t)\rho_{sym}(x^{'},t^{'})>_{0}  =  2\left[    \frac{i}{2\pi} \frac{ \frac{ \pi }{\beta v_F } }{\sinh( \frac{ \pi }{\beta v_F } (x-x^{'}-v_F(t-t^{'}) ) )  } \right]^2
\end{align}
and
\begin{align}
 <T \mbox{ }&\rho_{asy}(x,t)\rho_{asy}(x^{'},t^{'})>_{0}
  \nonumber \\&= \frac{\substack{-((4 v_{F}^{2} + \Gamma^{2})^{2}-32 v_{F}^{2} \Gamma^{2} \theta(x))((4 v_{F}^{2} + \Gamma^{2})^{2}-32 v_{F}^{2} \Gamma^{2} \theta(x^{'}))-64(-4 v_{F}^{3} \Gamma + v_{F} \Gamma^{3})^{2} \cos(\frac{- e V_{b}(x-x^{'}-v_F(t-t^{'}) )}{v_{F}})\mbox{ }\theta(x)\theta(x^{'})}}{\substack{2 v_{F}^{2} \beta^{2}(4 v_{F}^{2} + \Gamma^{2})^{4}\sinh( \frac{ \pi }{\beta v_F } (x-x^{'}-v_F(t-t^{'}) ) )^{2} }}
\end{align}
Let us consider the cases $ x > 0, x^{'} > 0 $ or  $ x > 0, x^{'} < 0 $ or $ x < 0, x^{'} > 0 $, the $\lambda=0$ term does not reproduce the correct terms for these cases as is evident from the presence of terms of the type $e^{ -(2\pi i)^2 \frac{1}{4}\int^{x} dy \mbox{ }    \int^{x^{'}} dy^{'} \mbox{ }
(<T\rho_{sym}(y,t)\rho_{sym}(y^{'},t^{'})>+<T\rho_{asy}(y,t)\rho_{asy}(y^{'},t^{'})> )}$. So for these cases $w_{0}=0$ and we can write Eq.\ref{rhoRC5} as,
\begin{align}
-<T\mbox{ }&\psi_R(x^{'},t^{'})\psi^{\dagger}_R(x_1,t_{1+})>  <T\mbox{ }\psi_R(x_1,t_1)\psi^{\dagger}_R(x,t)> \mbox{         } = \mbox{          }\nonumber \\
&w_{11} \mbox{    } \substack{( -2\pi i\int^{x} dy \mbox{ }
\frac{1}{2}  2\left[  \frac{i}{2\pi} \frac{ \frac{ \pi }{\beta v_F } }{\sinh( \frac{ \pi }{\beta v_F } (x_1-y-v_F(t_1-t) ) )  } \right]^2}
 \nonumber \\&\substack{+ 2\pi i \frac{1}{4}\int^{x^{'}} dy^{'} \mbox{ }
 (  2\left[    \frac{i}{2\pi} \frac{ \frac{ \pi }{\beta v_F } }{\sinh( \frac{ \pi }{\beta v_F } (x_1-y^{'}-v_F(t_1-t^{'}) ) )  } \right]^2
 +<\rho_{asy}(x_1,t_1)\rho_{asy}(y^{'},t^{'})>) )}\nonumber \\
 &  \mbox{    }  e^{ \frac{1}{2}(2\pi i)^2 \int^{x} dy \mbox{ }\int^{x} dy^{'} \mbox{ }
  2\left[    \frac{i}{2\pi} \frac{ \frac{ \pi }{\beta v_F } }{\sinh( \frac{ \pi }{\beta v_F } (y-y^{'} ) )  } \right]^2
   } e^{ \frac{1}{2} (2\pi i)^2 \frac{1}{4}
  \int^{x^{'}} dy^{'} \mbox{ } \int^{x^{'}} dy \mbox{ }
   (
   2\left[    \frac{i}{2\pi} \frac{ \frac{ \pi }{\beta v_F } }{\sinh( \frac{ \pi }{\beta v_F } (y^{'}-y  ) )  } \right]^2
   + < \rho_{asy}(y^{'},t^{'})\rho_{asy}(y,t^{'}) > )
  }\nonumber \\
  &\mbox{    }  e^{ -(2\pi i)^2 \frac{1}{2}\int^{x} dy \mbox{ }  \int^{x^{'}} dy^{'} \mbox{ }
      2\left[    \frac{i}{2\pi} \frac{ \frac{ \pi }{\beta v_F } }{\sinh( \frac{ \pi }{\beta v_F } (y-y^{'}-v_F(t-t^{'}) ) )  } \right]^2
     }\nonumber \\
   &  w_{21}\mbox{        }
\substack{( -2\pi i \frac{1}{4}\int^{x} dy \mbox{ }   (
 2\left[    \frac{i}{2\pi} \frac{ \frac{ \pi }{\beta v_F } }{\sinh( \frac{ \pi }{\beta v_F } (x_1-y-v_F(t_1-t) ) )  } \right]^2
+
 < \rho_{asy}(x_1,t_1)\rho_{asy}(y,t) >}
  \nonumber \\&\substack{+ 2 \pi i \frac{1}{2} \int^{x^{'}}dy^{'} 2\left[    \frac{i}{2\pi} \frac{ \frac{ \pi }{\beta v_F } }{\sinh( \frac{ \pi }{\beta v_F } (x_1-y^{'}-v_F(t_1-t^{'}) ) )  } \right]^2)}\nonumber \\
 & e^{ \frac{1}{2}(2\pi i)^2 \frac{1}{4}\int^{x} dy \mbox{ }\int^{x} dy^{'}
 \mbox{ }   (  2\left[    \frac{i}{2\pi} \frac{ \frac{ \pi }{\beta v_F } }{\sinh( \frac{ \pi }{\beta v_F } (y-y^{'}) )  } \right]^2
 + < \rho_{asy}(y,t)\rho_{asy}(y^{'},t)>)  } e^{ \frac{1}{2} ( 2\pi i)^2 \int^{x^{'}} dy^{'} \mbox{ } \int^{x^{'}} dy\mbox{ }
  2\left[    \frac{i}{2\pi} \frac{ \frac{ \pi }{\beta v_F } }{\sinh( \frac{ \pi }{\beta v_F } (y^{'}-y) )  } \right]^2
  }\nonumber \\
  &\mbox{    }    e^{ -(2\pi i)^2 \frac{1}{2} \int^{x} dy \mbox{ }   \int^{x^{'}} dy^{'} \mbox{ }
   2\left[    \frac{i}{2\pi} \frac{ \frac{ \pi }{\beta v_F } }{\sinh( \frac{ \pi }{\beta v_F } (y-y^{'}-v_F(t-t^{'}) ) )  } \right]^2
    }
\label{rhoRC8}
\end{align}
We include only the most singular terms of the expression in the RHS of the above equation.
Then Eq.\ref{rhoRC8} reduces to,
\begin{align}
-<T\mbox{ }&\psi_R(x^{'},t^{'})\psi^{\dagger}_R(x_1,t_{1+})>  <T\mbox{ }\psi_R(x_1,t_1)\psi^{\dagger}_R(x,t)> \mbox{         } = \mbox{          }\nonumber \\
&-\frac{i}{2\pi} \frac{\frac{\pi}{\beta v_{F}}}{\sinh( \frac{ \pi }{\beta v_F } ( x^{'}-x_{1}-v_F(t^{'}-t_{1}) ) )}
\mbox{         }\bigg(  U(t_1,t^{'}) \left[1
-      \theta(x_1)    \mbox{          }   \frac{  2 \Gamma^2
}{\Gamma ^2 +4 v_F^2}\right]\mbox{          }    \left[1
-      \theta(x^{'})    \mbox{          }   \frac{  2 \Gamma^2
}{\Gamma ^2 +4 v_F^2}\right]
  \nonumber \\ &+ \left( \frac{\Gamma }{v_F} \mbox{          } \frac{(2 v_F)^2
}{\Gamma ^2 +4 v_F^2}\right)^2 \mbox{          }
  \theta(x_1)     \theta(x^{'} )   \mbox{          } U(t_1,t_1 - \frac{x_1}{v_F})   \mbox{     }
   U(t^{'} - \frac{x^{'}}{v_F},t^{'})   \bigg)\nonumber \\
   & \mbox{ }\frac{i}{2\pi} \frac{\frac{\pi}{\beta v_{F}}}{\sinh( \frac{ \pi }{\beta v_F } ( x_{1}-x-v_F(t_{1}-t) ) )}
\mbox{       }
\bigg(  U(t,t_1) \left[1
-      \theta(x)    \mbox{          }   \frac{  2 \Gamma^2
}{\Gamma ^2 +4 v_F^2}\right]\mbox{          }    \left[1
-      \theta(x_1)    \mbox{          }   \frac{  2 \Gamma^2
}{\Gamma ^2 +4 v_F^2}\right]
  \nonumber \\ &+ \left( \frac{\Gamma }{v_F} \mbox{          } \frac{(2 v_F)^2
}{\Gamma ^2 +4 v_F^2}\right)^2 \mbox{          }
  \theta(x)     \theta(x_1)   \mbox{          } U(t,t - \frac{x}{v_F})   \mbox{     }
   U(t_1 - \frac{x_1}{v_F},t_1)   \bigg)\nonumber \\
   \mbox{         }& = \mbox{          }
  w_{11} \mbox{    } \frac{i}{4 \beta  v_F}\mbox{       }
   \left(\substack{2 \coth \left(\frac{\pi  (v_F(t-t_1)-x+x_1)}{\beta  v_F}\right)-\frac{2 \left(\left(\Gamma ^2+4 v_F^2\right)^2-16 \Gamma ^2 v_F^2 (-2 \theta (x_1,x^{'})+\theta (x_1)+\theta (x^{'}))\right)
 }{\left(\Gamma ^2+4 v_F^2\right)^2} \coth \left(\frac{\pi  (x_1-x^{'} -v_F(t_1-t^{'}))}{\beta  v_F}\right) }\right)\nonumber \\
 &\mbox{    }  e^{\Delta
   } e^{ \zeta(x^{'})
  } \mbox{    } \text{csch}\left(\frac{\pi }{\beta  v_F} (x-x^{'}-v_F(t-t^{'}))\right)\nonumber \\
  &+
w_{21}\mbox{        }\left(\substack{
( -2\pi i )\frac{1}{4}  \mbox{ }
 \frac{\left(\frac{32 \Gamma ^2 v_F^2 (-2 \theta (x,x_1)+\theta (x)+\theta (x_1))}{\left(\Gamma ^2+4 v_F^2\right)^2}-2\right) }{2 \pi  \beta  v_F}\mbox{ }
 \coth \left(\frac{\pi  (v_F(t-t_1)-x+x_1)}{\beta  v_F}\right)-\frac{i}{2 v_{F}\beta} \coth\left(\frac{\pi  (v_F(t^{'}-t_1)-x^{'}+x_1)}{\beta  v_F}\right)}\right)\nonumber \\
 & \mbox{     }
e^{ \zeta(x) } e^{ \Delta
  }\mbox{    } \text{csch}\left(\frac{\pi }{\beta  v_F} (x-x^{'}-v_F(t-t^{'}))\right)
\end{align}
where $\zeta(x) \mbox{ } = \mbox{ } \frac{1}{2}(2\pi i)^2 \frac{1}{4}\int^{x} dy \mbox{ }\int^{x} dy^{'}
 \mbox{ }   (  2\left[    \frac{i}{2\pi} \frac{ \frac{ \pi }{\beta v_F } }{\sinh( \frac{ \pi }{\beta v_F } (y-y^{'}) )  } \right]^2
 + < \rho_{asy}(y,t)\rho_{asy}(y^{'},t)>)$
 and
 $\Delta \mbox{ } = \mbox{ } \frac{1}{2} ( 2\pi i)^2 \int^{x^{'}} dy^{'} \mbox{ } \int^{x^{'}} dy\mbox{ }
  2\left[    \frac{i}{2\pi} \frac{ \frac{ \pi }{\beta v_F } }{\sinh( \frac{ \pi }{\beta v_F } (y^{'}-y) )  } \right]^2$. It is easy to appropriately fix the prefactors $w_{11}$ and $w_{21}$ such that the correct Wick's theorem expression for the four-point function is obtained. For the case $ x,x^{'},x_1 < 0$ Eq.\ref{rhoRC5} reduces to
\begin{align}
-&\frac{i}{2\pi} \frac{\frac{\pi}{\beta v_{F}}}{\sinh( \frac{ \pi }{\beta v_F } ( x^{'}-x_{1}-v_F(t^{'}-t_{1}) ) )}
\mbox{         }
  U(t,t^{'}) \mbox{          }
\mbox{ }\frac{i}{2\pi} \frac{\frac{\pi}{\beta v_{F}}}{\sinh( \frac{ \pi }{\beta v_F } ( x_{1}-x-v_F(t_{1}-t) ) )}\nonumber \\
\mbox{         } = \mbox{          }&
\frac{ 2\pi i \mbox{ } w_{0} }{4 \pi  \beta  v_F }\mbox{    }  (  \coth \left(\frac{\pi  (x_1-x + v_F(t-t_1))}{\beta v_F}\right)
  - \coth \left(\frac{\pi  (x_1-x^{'} -v_F(t_1-t^{'}))}{\beta  v_F}\right)   )\mbox{ } \mbox{  }
  \mbox{  }\nonumber \\& \sinh \left(\frac{\pi  \delta }{\beta  v_F }\right)
\mbox{         }
\text{csch}\left(\frac{\pi  (x-x^{'}-v_F(t-t^{'}))}{\beta v_F}\right)\nonumber \\
& +  w_{11} \mbox{    } \frac{i}{4 \beta  v_F}\mbox{       }
   \left(2 \coth \left(\frac{\pi  (v_F(t-t_1)-x+x_1)}{\beta  v_F}\right)-2 \coth \left(\frac{\pi  (x_1-x^{'} -v_F(t_1-t^{'}))}{\beta  v_F}\right) \right)\nonumber \\
&  \mbox{    }  e^{\Delta
   } e^{ \zeta_{<}
  } \mbox{    } \text{csch}\left(\frac{\pi }{\beta  v_F} (x-x^{'}-v_F(t-t^{'}))\right)\nonumber \\
&  +
w_{21}\mbox{        }\left(\frac{i}{2 v_{F}\beta} \mbox{ }
 \coth \left(\frac{\pi  (v_F(t-t_1)-x+x_1)}{\beta  v_F}\right)-\frac{i}{2 v_{F}\beta} \coth\left(\frac{\pi  (v_F(t^{'}-t_1)-x^{'}+x_1)}{\beta  v_F}\right)\right)\nonumber \\
 & \mbox{     }
e^{ \zeta_{<} } e^{ \Delta
  }\mbox{    } \text{csch}\left(\frac{\pi }{\beta  v_F} (x-x^{'}-v_F(t-t^{'}))\right)
\end{align}
and for $x,x^{'} < 0$, $x_{1} > 0$ Eq.\ref{rhoRC5} becomes
\begin{align}
-\mbox{         }
  &U(t,t^{'})\gamma_1
\mbox{ }\frac{i}{2\pi} \frac{\frac{\pi}{\beta v_{F}}}{\sinh( \frac{ \pi }{\beta v_F } ( x^{'}-x_{1}-v_F(t^{'}-t_{1}) ) )}
\frac{i}{2\pi} \frac{\frac{\pi}{\beta v_{F}}}{\sinh( \frac{ \pi }{\beta v_F } ( x_{1}-x-v_F(t_{1}-t) ) )}
\mbox{       } \nonumber \\
\mbox{         } = \mbox{          }&
w_{0} \mbox{    }  2\pi i\frac{1}{4} \frac{ \gamma_1}{\beta ^2 v_F^2}\frac{\beta v_F
}{\pi } (  \coth \left(\frac{\pi }{\beta v_{F}} (x_{1}-x+ v_F(t-t_1))\right)
- \coth \left(\frac{\pi }{\beta v_{F}} (x_{1}-x^{'}+ v_F(t^{'}-t_1))\right)  )
 \mbox{ }\nonumber \\&
\sinh \left(\frac{\pi  \delta }{\beta  v_F }\right)
\mbox{ }
\text{csch}\left(\frac{\pi  (x-x^{'} - v_F(t-t^{'}))}{\beta v_F}\right)\nonumber \\
&+   \frac{i}{2 \beta  v_F}\mbox{       }
   \bigg( (w_{11} + w_{21}\mbox{        }\gamma_1) \mbox{    } \coth \left(\frac{\pi}{\beta  v_F}  (v_F(t-t_1)-x+x_1)\right)\nonumber \\&-
   ( w_{11} \mbox{    }\gamma_1 + w_{21})\coth \left(\frac{\pi }{\beta  v_F} (x_1-x^{'} -v_F(t_1-t^{'}))\right) \bigg)
 \mbox{    }  e^{\Delta
   + \zeta_{<}  } \mbox{    } \text{csch}\left(\frac{\pi }{\beta  v_F} (x-x^{'}-v_F(t-t^{'}))\right)
   \label{CRRll}
\end{align}
where $\gamma_{1} = \frac{(\Gamma^{2}-4 v_{F}^{2})^{2}}{(\Gamma^{2}+4 v_{F}^{2})^{2}}$. 
Hence for $x,x^{'} < 0$ it appears that both $\lambda=0$ and $\lambda=1$ terms can be used to obtain the four-point functions with the only constraint on the prefactors being $w_{11} = w_{21}$ which is evident from Eq.\ref{CRRll}. But by evaluating $<T\mbox{ }\rho_L(-x_1,t_1)\psi^{\dagger}_R(x,t)\psi_R(x^{'},t^{'})> $ we'll see that only the conventional ($\lambda=0$) choice is the consistent one for $x, x^{'} < 0$ . We can write,
\begin{align}
<T\mbox{ }&\rho_L(-x_1,t_1)\psi^{\dagger}_R(x,t)\psi_R(x^{'},t^{'})>\nonumber \\&
=\mbox{ }-<T\mbox{ }\psi_R(x^{'},t^{'})\psi^{\dagger}_L(-x_1,t_1)>   <T\mbox{ }\psi_L(-x_1,t_{1})\psi^{\dagger}_R(x,t)> \nonumber \\ \mbox{ }&=\mbox{ } q_{0}\mbox{    } <\rho_L(-x_1,t_1) e^{ -2\pi i\int^{x} dy \mbox{ } \rho_R(y,t) } e^{ 2\pi i\int^{x^{'}} dy^{'} \mbox{ } \rho_R(y^{'},t^{'}) } > \nonumber \\
&\mbox{ }\mbox{ }\mbox{ }\mbox{ }+ q_{11}<\rho_L(-x_1,t_1) e^{ -2\pi i\int^{x} dy \mbox{ } (\rho_R(y,t)+\rho_L(-y,t)) } e^{ 2\pi i\int^{x^{'}} dy^{'} \mbox{ } \rho_R(y^{'},t^{'}) } > \nonumber \\
&\mbox{ }\mbox{ }\mbox{ }\mbox{ }+ q_{21} \mbox{    } <\rho_L(-x_1,t_1) e^{ -2\pi i\int^{x} dy \mbox{ }  \rho_R(y,t) } e^{ 2\pi i\int^{x^{'}} dy^{'} \mbox{ } (\rho_R(y^{'},t^{'})+\rho_L(-y^{'},t^{'})) } >
\label{rhorlc}
\end{align}
Following a similar procedure as shown above we can conclude that for the cases $ x > 0, x^{'} > 0 $ or  $ x > 0, x^{'} < 0 $ or $ x < 0, x^{'} > 0 $, the $\lambda=0$ term will not reproduce the correct form of the concerned four-point function. Considering only the most singular terms in Eq.\ref{rhorlc} we get,
\begin{align}
&\substack{-\frac{i}{2\pi} \frac{\frac{\pi}{\beta v_F}}{\sinh(\frac{\pi}{\beta v_F} ( x^{'} - x_1 - v_F(t^{'}-t_1) ))}
\mbox{     } \left(-U(t^{'}- \frac{x^{'}}{v_F},t^{'}) \left[ 1 - \theta(x_1) \frac{2\Gamma^2}{\Gamma^2 + 4v_F^2}\right] \theta(x^{'})
 + U(t_1 - \frac{x_1}{v_F},t^{'}) \left[ 1 - \theta(x^{'}) \frac{2\Gamma^2}{\Gamma^2 + 4v_F^2}\right] \theta(x_1) \right) i \frac{\Gamma}{v_F}\mbox{ } \frac{(2v_F)^2}{\Gamma^2 + 4 v_F^2}\nonumber \\
 \mbox{  }   \frac{i}{2\pi} \frac{\frac{\pi}{\beta v_F}}{\sinh(\frac{\pi}{\beta v_F} ( x - x_1 - v_F(t-t_1) ))}
\mbox{ }  \left(-U(t,t - \frac{x}{v_F}) \left[ 1 - \theta(x_1) \frac{2\Gamma^2}{\Gamma^2 + 4v_F^2}\right] \theta(x)
 + U(t,t_1 - \frac{x_1}{v_F}) \left[ 1 - \theta(x) \frac{2\Gamma^2}{\Gamma^2 + 4v_F^2}\right] \theta(x_1) \right) i \frac{\Gamma}{v_F}\mbox{ } \frac{(2v_F)^2}{\Gamma^2 + 4 v_F^2}} \\
 &\substack{\mbox{         } = \mbox{          }
  q_{11} \mbox{    } ( \frac{i }{2 \beta v_{F}}\coth \left(\frac{\pi
  }{\beta v_{F}}( x_1-x  + v_F(t-t_1)) \right)
 -\frac{8 i \pi  \Gamma ^2}{\beta ^2 \left(\Gamma ^2+4 v_{F}^2\right)^2} \mbox{ } \theta(-x_1 x^{'})
\mbox{ } \frac{\beta  v_{F}}{\pi } \coth \left(\frac{\pi
}{\beta  v_{F}}(x_1-x^{'} - v_{F} (t_{1}-t^{'})) \right) )
 \mbox{    } \\ e^{\Delta
   } e^{ \zeta(x^{'})
  } \mbox{    } \text{csch}\left(\frac{\pi  }{\beta  v_{F}}   (x-x^{'} -v_F(t-t^{'}))  \right)}\nonumber \\
  &\substack{\mbox{ }\mbox{ }\mbox{ }\mbox{ }+
q_{21}\mbox{        }
( - \frac{i }{2 \beta v_{F}}\coth \left(\frac{\pi
  }{\beta v_{F}}( x_1-x^{'}  + v_F(t^{'}-t_1)) \right)
 + \frac{8 i \pi  \Gamma ^2}{\beta ^2 \left(\Gamma ^2+4 v_{F}^2\right)^2} \mbox{  } \theta(-x_1 x)
\mbox{ } \frac{\beta  v_{F}}{\pi } \coth \left(\frac{\pi
}{\beta  v_{F}}(x_1-x - v_{F} (t_{1}-t)) \right) )
\mbox{     }  \nonumber \\e^{ \Delta
  }
e^{ \zeta(x) }
 \mbox{    } \text{csch}\left(\frac{\pi  }{\beta  v_{F}}   (x-x^{'} -v_F(t-t^{'}))  \right)}
\end{align}
Solving for the prefactors so that the above equation holds gives us,
\begin{align}
q_{11}= - \frac{ i\eta e^{-\Delta - \zeta(x^{'})}\left(\Gamma ^2+4 v_{F}^2\right)^2 (  \left(\Gamma ^2+4 v_{F}^2\right)^2-16 \Gamma ^2  v_F^2
\theta (-x x_{1}) ) }{2 \beta v_{F} ( \left(\Gamma ^2+4 v_{F}^2\right)^4-256   \Gamma^4 v_{F}^4   \theta (-x x_{1}) \theta (-x_{1} x^{'}) ) }
\end{align}
\begin{align}
q_{21}=-\frac{i \eta  e^{-\Delta - \zeta(x)} \left(\Gamma ^2+4 v_{F}^2\right)^2 \left(\left(\Gamma^2+4 v_{F}^2\right)^2-16 \Gamma ^2 v_{F}^2 \theta (-x_{1} x^{'})\right)}{2 \beta  v_{F} \left(\left(\Gamma^2+4 v_{F}^2\right)^4-256 \Gamma^4 v_{F}^4 \theta (-x x_{1},-x_{1} x^{'})\right)}
\end{align}
where,
\begin{align}
\eta = &\substack{-\frac{16 \Gamma^{2} v_{F}^{2}}{(\Gamma^{2} + 4 v_{F}^{2})^{2}}\left(-U(t^{'}- \frac{x^{'}}{v_F},t^{'}) \left[ 1 - \theta(x_1) \frac{2\Gamma^2}{\Gamma^2 + 4v_F^2}\right] \theta(x^{'})
 + U(t_1 - \frac{x_1}{v_F},t^{'}) \left[ 1 - \theta(x^{'}) \frac{2\Gamma^2}{\Gamma^2 + 4v_F^2}\right] \theta(x_1) \right)}\nonumber \\
 &\substack{\left(-U(t,t - \frac{x}{v_F}) \left[ 1 - \theta(x_1) \frac{2\Gamma^2}{\Gamma^2 + 4v_F^2}\right] \theta(x)
 + U(t,t_1 - \frac{x_1}{v_F}) \left[ 1 - \theta(x) \frac{2\Gamma^2}{\Gamma^2 + 4v_F^2}\right] \theta(x_1) \right)}
 \label{eta}
\end{align}
So using only the terms with $\lambda=1$ in the exponent we obtain the correct four-point function as well as the correct two-point functions for these cases. Now let us consider the case $x<0, x^{'}<0$ in Eq.\ref{rhorlc} and including only the most singular terms we get,
\begin{align}
&\substack{-\frac{i}{2\pi} \frac{\frac{\pi}{\beta v_F}}{\sinh(\frac{\pi}{\beta v_F} ( x^{'} - x_1 - v_F(t^{'}-t_1) ))}
\mbox{     } \left(-U(t^{'}- \frac{x^{'}}{v_F},t^{'}) \left[ 1 - \theta(x_1) \frac{2\Gamma^2}{\Gamma^2 + 4v_F^2}\right] \theta(x^{'})
 + U(t_1 - \frac{x_1}{v_F},t^{'}) \left[ 1 - \theta(x^{'}) \frac{2\Gamma^2}{\Gamma^2 + 4v_F^2}\right] \theta(x_1) \right) i \frac{\Gamma}{v_F}\mbox{ } \frac{(2v_F)^2}{\Gamma^2 + 4 v_F^2}\nonumber \\
 \mbox{  }   \frac{i}{2\pi} \frac{\frac{\pi}{\beta v_F}}{\sinh(\frac{\pi}{\beta v_F} ( x - x_1 - v_F(t-t_1) ))}
\mbox{ }  \left(-U(t,t - \frac{x}{v_F}) \left[ 1 - \theta(x_1) \frac{2\Gamma^2}{\Gamma^2 + 4v_F^2}\right] \theta(x)
 + U(t,t_1 - \frac{x_1}{v_F}) \left[ 1 - \theta(x) \frac{2\Gamma^2}{\Gamma^2 + 4v_F^2}\right] \theta(x_1) \right) i \frac{\Gamma}{v_F}\mbox{ } \frac{(2v_F)^2}{\Gamma^2 + 4 v_F^2}}\nonumber \\
 &\substack{\mbox{         } = \mbox{          }
  q_{11} \mbox{    } ( \frac{i }{2 \beta v_{F}}\coth \left(\frac{\pi
  }{\beta v_{F}}( x_1-x  + v_F(t-t_1)) \right)
 -\frac{8 i \pi  \Gamma ^2}{\beta ^2 \left(\Gamma ^2+4 v_{F}^2\right)^2} \mbox{ } \theta(-x_1 x^{'})
\mbox{ } \frac{\beta  v_{F}}{\pi } \coth \left(\frac{\pi
}{\beta  v_{F}}(x_1-x^{'} - v_{F} (t_{1}-t^{'})) \right) )
 \mbox{    } \\ e^{\Delta
   } e^{ \zeta(x^{'})
  } \mbox{    } \text{csch}\left(\frac{\pi  }{\beta  v_{F}}   (x-x^{'} -v_F(t-t^{'}))  \right)}\nonumber \\
  &\substack{\mbox{ }\mbox{ }\mbox{ }\mbox{ }+
q_{21}\mbox{        }
( - \frac{i }{2 \beta v_{F}}\coth \left(\frac{\pi
  }{\beta v_{F}}( x_1-x^{'}  + v_F(t^{'}-t_1)) \right)
 + \frac{8 i \pi  \Gamma ^2}{\beta ^2 \left(\Gamma ^2+4 v_{F}^2\right)^2} \mbox{  } \theta(-x_1 x)
\mbox{ } \frac{\beta  v_{F}}{\pi } \coth \left(\frac{\pi
}{\beta  v_{F}}(x_1-x - v_{F} (t_{1}-t)) \right) )
\mbox{     }  \\ e^{ \Delta
  }
e^{ \zeta(x) }
 \mbox{    } \text{csch}\left(\frac{\pi  }{\beta  v_{F}}   (x-x^{'} -v_F(t-t^{'}))  \right)}\nonumber \\
 &\substack{\mbox{ }\mbox{ }\mbox{ }\mbox{ }+q_{0} \mbox{    } \frac{8 i \Gamma ^2 v_F \theta (x_1)
 \left(\coth \left(\frac{\pi  }{\beta  v_F} (x_1-x - v_F(t_1-t))\right)
 -\coth \left(\frac{\pi  }{\beta  v_F}(x_1-x^{'}-v_F(t_1-t^{'}))\right)\right)}{\beta  \left(\Gamma ^2+4 v_F^2\right)^2}
  \mbox{    } \sinh \left(\frac{\pi  \delta }{\beta  v_F }\right)\mbox{     }
\text{csch}\left(\frac{\pi }{\beta v_{F}} (x-x^{'} - v_F(t-t^{'}))\right)}
\end{align}
The only relevant case here is $x<0, x^{'}<0, x_{1}>0$ as the Wick's theorem result is zero for $x<0, x^{'}<0, x_{1}>0$. A subtle point to note is that in both the four-point functions of interest $<T\mbox{ }\rho_L(-x_1,t_1)\psi^{\dagger}_R(x,t)\psi_R(x^{'},t^{'})>$ and $<T\mbox{ }\rho_R(x_1,t_1)\psi^{\dagger}_R(x,t)\psi_R(x^{'},t^{'})>$, we write the term $\psi^{\dagger}_R(x,t)\psi_R(x^{'},t^{'})$ in bosonized form which is the same in both cases with the only difference being the presence of the $\rho_L(-x_1,t_1)$ or $\rho_R(x_1,t_1)$ term when evaluating the expectation value. This means that the prefactors in both the cases i.e. the $w$'s and the $q$'s should have the same qualitative properties in order to match the Wick's theorem result for the four-point functions. We have already obtained the constraint $w_{11} = w_{21}$ for $x,x^{'} < 0$. So it follows that $q_{11} = q_{21}$ for $x<0, x^{'}<0, x_{1}>0$ and solving for the prefactors to obtain the correct Wick's theorem result we get $q_{11} = q_{21} = 0$ and $q_{0}\mbox{     } = \mbox{           }
 -\frac{i \eta  \left(\Gamma ^2+4 v_{F}^2\right)^2 \text{csch}\left(\frac{\pi  \delta }{\beta v_{F}}\right)}{32 \beta  \Gamma ^2 v_{F}^3}$, where $\eta$ is as defined in Eq.\ref{eta}. This means that the unconventional bosonization choice does not play any role this case as its prefactors are identically zero and it entirely drops out from the calculation. The implication here is that the conventional bosonization procedure (ie. without the anomalous term in the exponent) is the only valid choice for writing the bosonized version of the $< T\mbox{ }\Psi_R(x<0,t) \Psi^{\dagger}_R(x^{'}<0,t^{'})>$ correlation function. A similar analysis shows that for $< T\mbox{ }\Psi_L(x>0,t) \Psi^{\dagger}_L(x^{'}>0,t^{'})>$ also only the conventional method of bosonization is valid. But for all the other cases it is necessary to include the unconventional anomalous term in the exponent to obtain the correct form of the Green functions.
 
 \section*{APPENDIX D: Universal power-law behaviour in NCBT Green's functions of strongly inhomogeneous Luttinger liquids in equilibrium}
\label{AppendixD}
\setcounter{equation}{0}
\renewcommand{\theequation}{D.\arabic{equation}}
The most singular parts of the full interacting Green's functions of a strongly inhomogeneous Luttinger liquid in equilibrium obtained using NCBT are shown in Eqs.14 and 15 of \cite{doi:10.1142/S0217751X18501749}. Substituting $x_{1} = x$, $x_{2} = x+\epsilon$ (to avoid infinities/zeroes) and $t_{1} = t_{2} = t$ in the RL same side ($x_{1}$ and $x_{2}$ on same side of the origin) Green's function, we obtain
\begin{align}
<\psi_{R}(x)\psi^{\dagger}_{L}(x)> \mbox{ }\sim\mbox{ } &x^{X}(-2 x - \epsilon)^{C-X}(-\epsilon)^{2 Q-\frac{Q}{C}} (x+\epsilon)^{X} (2 x + \epsilon)^{-1+C-X}(x^{2 C} + (x+\epsilon)^{2 C})\nonumber \\
\sim \mbox{ }& x^{-g}
\end{align}
where $g = \frac{v_{F}}{v_{h}}$ is just the Luttinger liquid interaction parameter.
The expressions for the anomalous exponents that appear in the NCBT Green functions are as follows,
\begin{equation*}
Q = \frac{(v_{h}-v_{F})^2}{8 v_{h} v_{F}} \mbox{ };\mbox{ }X = \frac{|R|^{2} (v_{h}-v_{F})(v_{h}+v_{F})}{8v_{h}(v_{h} - |R|^{2}(v_{h}-v_{F}))} \mbox{ }; \mbox{ }C = \frac{v_{h}-v_{F}}{4v_{h}}
\end{equation*}
where $v_{F}$ is the Fermi velocity, $v_{h}$ is the holon velocity and $|R|$ is the reflection amplitude. Hence in the limit $x_{1} \rightarrow x_{2}$ the power law exponents turn out to be universal (i.e. independent of impurity strength) although the exponents in the general expresssion for the Green functions do depend on the impurity strength , hence termed anomalous exponents.  
For the RR same side Green's function we obtain,
\begin{align}
<\psi_{R}(x)\psi^{\dagger}_{R}(x)> \mbox{ }\sim \mbox{ } &4^{X}(-2 x-\epsilon)^{-X}(-\epsilon)^{-1-Q}(\epsilon)^{-Q}(x(x+\epsilon))^{X}(2 x + \epsilon)^{-X}\nonumber \\
\sim \mbox{ }& x^{0} 
\end{align}
The $x$-dependence drops out in the RR and LL cases.
The impurity dependence in the exponent drops out in one case and the final power-law exponent is universal as it depends only on the interaction parameter and in the other case the overall exponent adds up to zero making it trivially universal. On the basis of these observations, it is reasonable to suspect that similar universal behaviour (i.e. impurity strength independence) may be expected even in a system driven out of equilibrium by the application of a bias. The equal space-time Green functions appear in the expression for the tunneling current as defined in Eq.\ref{eqitun}, hence we expect to observe universal scaling behaviour in the tunneling transport properties. This important work is relegated to a future publication.
\newpage
\section*{References}
\bibliographystyle{iopart-num}
\bibliography{ref}
\end{document}